\documentclass[12pt]{article}
\usepackage{amsfonts,cite}

\newcommand{\be}{\begin{equation}}
\newcommand{\ee}{\end{equation}}
\newcommand{\bea}{\begin{eqnarray}}
\newcommand{\eea}{\end{eqnarray}}
\newcommand{\ba}{\begin{array}}
\newcommand{\ea}{\end{array}}

\newcommand{\al}{\alpha}
\newcommand{\ga}{\gamma}
\newcommand{\Ga}{\Gamma}
\newcommand{\bet}{\beta}
\newcommand{\ka}{\kappa}
\newcommand{\de}{\delta}
\newcommand{\ep}{\epsilon}

\newcommand{\la}{\lambda}

\newcommand{\Om}{\Omega}
\newcommand{\ze}{\zeta}

\newcommand{\La}{\Lambda}

\newcommand{\Ups}{\Upsilon}

\newcommand{\Z}{\mathbb{Z}}
\newcommand{\R}{\mathbb{R}}

\newcommand{\D}{{\rm d}}
\newcommand{\ip}{{\rm i}}
\newcommand{\id}{\hbox{1\kern-.27em l}}
\newcommand{\sid}{\hbox{\scriptsize1\kern-.27em l}}
\newcommand{\tr}{{\rm tr}}
\newcommand{\pa}{\partial}
\newcommand{\rar}{\rightarrow}
\newcommand{\non}{\nonumber}
\newcommand{\we}{\kern-.1em\wedge\kern-.1em}
\newcommand{\scal}{\kern-.13em\cdot\kern-.13em}
\newcommand{\tF}{\tilde{F}}
\newcommand{\tA}{\tilde{A}}
\newcommand{\tK}{\tilde{K}}
\newcommand{\bLa}{{\bar{\La}}}
\newcommand{\bE}{{\bar{E}}}
\newcommand{\half}{\mbox{$\frac{1}{2}$}}
\newcommand{\third}{\mbox{$\frac{1}{3}$}}
\newcommand{\fourth}{\mbox{$\frac{1}{4}$}}
\newcommand{\sixth}{\mbox{$\frac{1}{6}$}}
\newcommand{\II}{I\kern-.09em I}

\begin{document}

\vspace*{-6mm}

\rightline{\vbox{\footnotesize
\hbox{DAMTP-1999-48}
\hbox{\tt hep-th/9904117}
}}

\vspace{3mm}

\begin{center}
{\Large\sf Supersymmetric brane actions from interpolating dualisations}

\vskip 5mm

Anders Westerberg\footnote{\tt A.Westerberg@damtp.cam.ac.uk}
and Niclas Wyllard\footnote{\tt N.Wyllard@damtp.cam.ac.uk} \vspace{5mm}\\
{\em DAMTP, University of Cambridge,\\ Silver Street, Cambridge CB3 9EW, UK}

\end{center}
 
\vskip 5mm
 
\begin{abstract}
\noindent
We explore the possibility of constructing $p$-brane world-volume actions with
the requirements of $\ka$-symmetry and gauge invariance as the only input. 
In the process, we develop a general framework which leads to actions
interpolating between Poincar\'e-dual descriptions of the world-volume
theories. The method does not require any restrictions on the on-shell
background configurations or on the dimensions of the branes. 
After some preliminary studies of low-dimensional cases we apply the method 
to the type {\II}B five-branes and, in particular, construct a $\ka$-symmetric 
action for the type {\II}B NS5-brane with a world-volume field content
reflecting the fact that the D1-, D3- and D5-branes can end on it. 
\end{abstract}

\setcounter{equation}{0}
\section{Introduction}

It is by now well established that extended objects play a fundamental r\^ole 
in the non-perturbative completion of the superstring theories. 
A few years ago this fact led to renewed interest in the
study of effective world-volume actions describing such objects. 
In particular, while actions for the fundamental strings and for the M-theory 
membrane~\cite{Bergshoeff:1987} have been known for a long time,
actions for the D-branes were constructed more recently
in refs~\cite{Leigh:1989,Cederwall:1996b,Cederwall:1996c,Aganagic:1996,
Bergshoeff:1996c}.
A further development was the observation that these branes can
equivalently be described by actions where the tension is generated
dynamically by a gauge-invariant world-volume $p{+}1$-form, $F_{p+1}$. 
These actions have the generic form
\begin{equation}
\label{action}
S = \int\D^{p+1}\xi\sqrt{-g}\,\la\,[1 + \Phi(\{F_{i}\}) - ({*}F_{p+1})^2]\,,
\end{equation}
where $\{F_{i}\}$ collectively denotes the world-volume field
strengths (excluding $F_{p+1}$), and $\la$ is a Lagrange 
multiplier for the constraint $1 +
\Phi({\{F_i\}}) - ({*}F_{p+1})^2 \approx 0$. The world-volume field strengths
(including $F_{p+1}$) have the schematic form 
$F= \D A - C - ``C\we F\mbox{''}$. Here $``C\we F\mbox{''}$ denotes corrections
to the canonical, minimally coupled form determined by the requirement of 
gauge invariance. The equation of motion for the tension gauge potential 
$A_p$ leads to $\la\,{*}F_{p+1} = {\rm const}$; this constant is identified 
with the tension. 

An appealing feature of the formulation (\ref{action}) is that, in all known 
cases, $\Phi$ 
is a polynomial function of its arguments. Notice also that there is no 
Wess--Zumino term in the above action; this term is instead incorporated 
in the $p{+}1$-form field strength $F_{p+1}$. Whenever convenient, one can 
return to the formulation without the Lagrange multiplier $\la$ and the
tension gauge potential $A_p$ by solving their equations
of motion, thereby regaining the Wess--Zumino term in its traditional form. 

Actions of the form (\ref{action}) were constructed for the M2-brane and the 
fundamental strings in refs~\cite{Townsend:1992,Bergshoeff:1992} and more 
recently for the D-branes in ref.~\cite{Bergshoeff:1998c}. 
For the type {\II}B branes it is known \cite{Schwarz:1995} that the
fundamental string (charge (1,0)) and the Dirichlet string (charge (0,1))
belong to a doublet of $(p,q)$ strings under the
non-perturbative SL(2,$\Z$) symmetry \cite{Hull:1995} of the
type {\II}B theory, and an action for these $(p,q)$ strings which is 
manifestly covariant under the SL(2,$\Z$) symmetry 
\cite{Townsend:1997,Cederwall:1997} has been constructed. The D3-brane
on the other hand is a singlet under SL(2,$\Z$). In
the action which makes this property manifest, two world-volume two-form 
field strengths
are required \cite{Cederwall:1998a}. In order to get the correct counting 
for the degrees of freedom, an auxiliary duality relation between these 
two fields has to be imposed at the level of the equations of motion. In
ref.~\cite{Cederwall:1998b} an action for the M5-brane was constructed which
circumvents the problems \cite{Witten:1996a} associated with the
self-dual world-volume three-form by implementing the self-duality
relation of this three-form as an auxiliary condition at the level of the 
equations of motion. Thus, in the last two cases duality relations had to be 
imposed to compensate for the fact that too many fields appear in the actions.
For the three-brane the introduction of extra fields was necessary to make a 
symmetry manifest, and for the M5-brane to circumvent a topological 
restriction. 
In this paper auxiliary duality relations will be a recurrent theme.

A crucial requirement for supersymmetric boson-fermion matching of the
above brane actions is that of $\ka$-symmetry, a local world-volume
symmetry for which the variation parameter $\ka$ is a target-space spinor 
which can be written as $\ka=P_{+}\ze=\half(\id+\Ga)\ze$, where $P_{\pm}$ are
mutually orthogonal projection operators, each reducing the number of 
independent components of a spinor by half. These properties translate 
into the requirements $\tr(\Ga)=0$, and $\Ga^2 = \id$.
A long-standing problem has been the construction of a $\ka$-symmetric
action for the type {\II}B five-brane. While it is known \cite{Callan:1991}
that the theory on the world-volume is described (on-shell) by a 
six-dimensional vector multiplet, there has been a debate in the literature 
about whether
this multiplet should be realised in the action as a two-form or as
a four-form field strength (these are Poincar\'e dual, and hence have 
the same number of degrees of freedom). Recently, a proposal based on 
T-duality considerations for the bosonic part of the action for the
type {\II}B NS5-brane action was put forward \cite{Eyras:1998}. 
One of the main results of the present paper is the construction of a
$\ka$-symmetric action for the type {\II}B NS5-brane in a general curved
supergravity background, and with a world-volume field content which reflects
the fact that the D1-, D3- and D5-branes can end on it. The bosonic part
of our action differs from the one given in ref.~\cite{Eyras:1998}; we will
comment on this fact in later sections.

In ref.~\cite{Townsend:1996a} it was observed that in order to relate the
action for the directly dimensionally reduced M2-brane to the action for the
D2-brane, which describes the same physical object, 
one has to perform a world-volume Poincar\'e dualisation of the one-form
on the world-volume of the dimensionally reduced membrane to a two-form,
which can be interpreted as the two-form field-strength 
on the D2-brane world-volume,
and vice versa. Such dualisations were shown to be required also in other
interconnections between $p$-brane actions \cite{Tseytlin:1996,Aganagic:1997}.
There are however limitations on the applicability of this method. 
The extension for $p\!>\!2$ to general backgrounds, where all form fields are 
non-constant, was recently accomplished in ref.~\cite{Kimura:1999a}. 
In the present paper we also work with general backgrounds. 
The method of refs~\cite{Tseytlin:1996,Aganagic:1997} runs into more serious 
problems when $p\!>\!4$, since one then encounters fifth-order algebraic 
equations. 

In this paper we propose a general method for constructing world-volume
actions with the requirements of $\ka$-symmetry and gauge invariance as the 
only input. To be able to apply the method one does not need to know even the 
bosonic part of the action beforehand. We work within the framework where the
brane actions take the form~(\ref{action}).
The method furthermore leads to a new way of handling Poincar\'e
dualisation, which in a certain sense circumvents the above mentioned problem. 
The dualisation is also more general in that it is not a
$\Z_2$ transformation; rather a whole set of interpolating theories
is constructed. The interpolating actions depend on
certain parameters which control the dualisation. For intermediate values of 
the parameters the world-volume fields are ``doubled,'' i.e., they
come in pairs with their Poincar\'e duals. 
In these formulations it can be arranged so that there is a world-volume
field corresponding to every possible brane-ending-on-another-brane
case~\cite{Strominger:1996,Townsend:1996b,Argurio:1997}, 
thus realising an equality among the various possible gauge
invariant world-volume field strengths on the branes. In order to obtain 
the correct number of degrees of freedom for intermediate values of the 
parameters auxiliary duality relations are imposed. As a byproduct of our 
results we show that the form of the usual D-brane actions is determined by 
super- and $\ka$-symmetry alone (supplemented by gauge invariance).
 
The fundamental requirements we impose are that the world-volume fields 
should be gauge invariant and satisfy Bianchi identities on the world-volume.
For the type {\II}B branes we also require our actions to be invariant
under the perturbative Peccei--Quinn symmetry (this requirement is closely 
related to the condition that the world-volume field strengths should satisfy 
Bianchi identities). All hitherto known supersymmetric brane actions
are invariant under this symmetry.

The paper is organised as follows. In the next section we illustrate the
method for the D2-brane in the type {\II}A theory. 
In section~\ref{IIBbranesect}
we discuss the strings and the D3-brane in the type {\II}B theory. 
The construction of the action for the type {\II}B NS5-brane is presented
in section~\ref{IIB5branesect}; this section also contains a discussion about
the D5-brane. 

Finally, in the appendices we describe our notation and conventions, list 
some properties of the ten-dimensional type {\II} supergravities in 
superspace relevant for our discussion, and give some details about
the method used to verify $\ka$-symmetry of our actions. 
In the latter context, we list a number of essential identities involving the 
world-volume field strengths. These identities should also be useful in
applications of our results.

\setcounter{equation}{0}
\section{An introductory example: the D2/M2-brane}

In this section we will describe the method in a simple, but non-trivial,
setting---the D2-brane in type {\II}A. This brane can also be described as
the dimensionally reduced M2-brane, the two actions being related via a 
Poincar\'e-duality transformation of the
world-volume fields \cite{Townsend:1996a}\footnote{In order to obtain the 
M2-brane action from the D2-brane action one has to perform a world-volume 
Poincar\'e-duality transformation followed by an S-duality transformation, 
which, given the relation between the eleventh dimension and the dilaton 
\cite{Townsend:1995a,Witten:1995a}, corresponds to lifting the action to 
eleven dimensions.}. 
This is, in fact, the way the $\ka$-symmetric action for the D2-brane was 
first constructed~\cite{Townsend:1996a}. 
The action for the dimensionally reduced M2-brane contains the world-volume
field $F^{\rm M2}_3 = \D A_2 - C_3 + B_2 \, F_1$ which generates the tension
(here $F_1 = \D A_0 - C_1$). 
Similarly, an action for the D2-brane where the tension is replaced by
the world-volume field $F^{\rm D2}_3 = \D A_2 - C_3 - C_1 \, F_2$
has been constructed~\cite{Bergshoeff:1998c} 
(here as usual $F_2 = \D A_1 - B_2$).

Before we continue we will briefly discuss some facts about the background
type {\II}A supergravity theory (more details can be found in 
appendix~\ref{sugraapp}). 
The gauge-invariant field strengths in the type {\II}A
theory which are relevant for the discussion in this section are
$R_2$, $H_3$ and $R_4$. These fields satisfy the Bianchi identities 
\be
\D R_2 = 0\,, \qquad \D H_3 = 0\,, \qquad \D R_4 = H_3\, R_2\,.
\ee
The first two relations are ``solved'' by $R_2 = \D C_1$ and $H_3 = \D B_2$, 
whereas there is an ambiguity in the definition of $R_4$; 
if one requires $R_4$ to satisfy the above Bianchi identity one arrives at 
the natural ``solution'' $R_4 = \D C_3 + x\,B_2 \, R_2 -
(1{-}x)\,C_1\, H_3 = \D (C_3 + x\,C_1\, B_2) - C_1\, H_3$. 
{}From the last equality we see that the parameter $x$ arises from an ambiguity
in the definition of $C_3$, the most general natural field redefinition
being $C_3 \rar C_3 + x\, C_1 \, B_2$. This innocent field redefinition will 
play a central role in the sequel. Different values of the
parameter $x$ describe the same physics, but, as we will see later, the
corresponding descriptions of the world-volume theories can be quite 
different. The background field strengths
are invariant under the gauge transformations 
\bea
\label{backgaug}
\de C_1 &=& \D L_0\,, \qquad\quad \de B_2 = \D L_1\,, \non\\
\de C_3 &=& \D L_2 - x\,R_2 \, L_1 - (1{-}x)\,H_3\, L_0\,.
\eea

We now turn to the construction of gauge-invariant world-volume field 
strengths satisfying Bianchi identities. 
The general form of these field strengths is 
$F = \D A - C - ``C\wedge F{\mbox{''}}$. Here $C$ collectively denotes the 
background potentials (both the RR and the NS-NS ones) and 
$``C\wedge F{\mbox{''}}$ denotes possible corrections to the minimally coupled 
form, required by gauge invariance. 
It is easy to see that $F_1 = \D A_0 - C_1$, and $F_2 = \D A_1 - B_2$
are invariant under the gauge transformations (\ref{backgaug})
together with
$\de A_0 = c + L_0$ and $\de A_1 = \D l_0 + L_1$ (here $c$ is a constant),
and that they satisfy the Bianchi identities 
\be
\label{BIF12}
\D F_1 = -R_2 \,, \qquad \D F_2 = -H_3\,.
\ee
Furthermore, using the above form of $R_4$, we see that the world-volume 
three-form field strength 
\be
F_3 = \D A_2 - C_3 + x\, B_2\, F_1 - (1{-}x)\, C_1\, F_2
\ee 
obeys the Bianchi identity 
\be
\label{BIF3}
\D F_3 = -R_4 + x\,F_1 \, H_3 - (1{-}x)\,F_2 \, R_2\,,
\ee
and is invariant under the above gauge transformations combined with 
$\de A_2 = \D l_1 + L_2 + x\,L_1\, F_1 + (1{-}x)\, L_0\, F_2$. 
Thus, on the world-volume the field redefinition in the background has
a more profound implication: it determines which world-volume field
strengths appear. We see that for $x=1$ we obtain the correct
tension field for the description of the dimensionally reduced M2-brane,
whereas for $x=0$ we instead find the D2-brane description. 
The limits $x=0$ and $x=1$ thus lead to sensible results, 
but what happens for other values of $x$? The answer to this question
is given below, where we construct an action which interpolates 
between the two limiting cases discussed at the beginning of the section.
The interpolation is controlled by the real parameter $x$ introduced
above. In order for the kinetic term to be positive semi-definite, it appears 
desirable to impose the physical restriction $x\in [0,1]$.
We make a general Ansatz for the action of the form
\be
\label{D2ansatz}
S = \int \D^{3}\xi\sqrt{-g}\,\la \left[1 +
\Phi(e^{\frac{3}{4}\phi}F_1,e^{-\frac{1}{2}\phi}F_2) - 
e^{\frac{2}{4}\,\phi}({*}F_3)^2\right]\,.
\ee
The dilaton dependence in this expression can be motivated by 
considering the dilaton-scaling of the supergravity constraints, which, 
via the basic $\ka$-variations, directly determine the dilaton-factor 
in front of each occurrence of the world-volume field strengths. We will 
from now on suppress the dilaton factors; they can be reinstated at any 
time by using the rules $F_1 \rar e^{\frac{3}{4}\phi}F_1$,
$F_2\rar e^{-\frac{1}{2}\phi}F_2$ and $F_3\rar e^{\frac{1}{4}\phi}F_3$.
The equation of motion for $A_2$ is $\D [\la\,{*}F_{3}]=0$, whereas the ones
for $A_0$ and $A_1$ are 
\bea
\D\Big[\la\{{*}\frac{\de\Phi}{\de F_1} + 2 x\, B_2 \,{*}F_3\}\Big] &=& 0\,,
\non \\
\D\Big[\la\{{*}\frac{\de\Phi}{\de F_2} - 2 (1{-}x)\,C_1\,{*}F_3\}\Big] &=& 0\,.
\eea
By using the Bianchi identities for $F_1$ and $F_2$ together with the
result $\D[\la\,{*}F_{3}]=0$ one can eliminate the
explicit appearance of the background field strengths: 
\bea
\D\Big[\la\{{*}\frac{\de\Phi}{\de F_1} - 2x\,F_2\,{*}F_3\}\Big] &=& 0\,,\non \\
\D\Big[\la\{{*}\frac{\de\Phi}{\de F_2} + 2 (1{-}x)\, F_1\, {*}F_3\}\Big] 
&=& 0\,.
\eea
It is thus consistent with the equations of motion and the Bianchi
identities to impose the ``duality'' relations
\bea
\label{D2dualrel}
-2x\,{*}F_3\,{*}F_2 &=& K_1 := \frac{\de\Phi}{\de F_1}\,,\non\\
2(1{-}x)\,{*}F_3\,{*}F_1 &=& K_2 := \frac{\de\Phi}{\de F_2}\,,
\eea
where $\Phi$ is yet to be determined. 
These relations reduce the number of degrees of freedom by half, and thus 
remedy the doubling of fields in the action. Auxiliary duality relations 
of this kind have been used before
in the literature \cite{Bergshoeff:1996e,Cederwall:1998a,Cederwall:1998b}. 
For the two limiting values $x=0$ and $x=1$ the duality 
relations~(\ref{D2dualrel}) become degenerate, so that we have the
same number of degrees of freedom for all values of $x$, namely that of 
a one-form $F_1$ (or, equivalently, of its Poincar\'e dual $F_2$). 

Turning next to the $\ka$-symmetry properties of the action~(\ref{D2ansatz}), 
it can be shown that the $\ka$-transformations of the world-volume fields are
\bea
\de_\ka g_{ij} &=& 2\,{E_{(i}}^a\,{E_{j)}}^B\ka^\al\,T_{\al Ba}\,, 
\quad
\de_\ka\phi= \ka^\al\pa_\al\phi \,, \non\\
\de_\ka F_1 &=& -i_{\ka}R_1\,, 
\hspace{23.5mm}
\de_{\ka}F_2 = -i_\ka H_3\,, \non\\
\de_\ka F_3 &=& -i_{\ka}R_4+x\,F_1\,i_{\ka}H_3-(1{-}x)\,F_2\,i_{\ka}R_2\,.
\eea
Notice the close correspondence between the variations of the world-volume
field strengths and their respective Bianchi identities given in 
eqs~(\ref{BIF12}) and (\ref{BIF3}).
This correspondence holds for all cases considered in this paper. 
We would now like to check whether the action~(\ref{D2ansatz}) is invariant
under these transformations. To this end, it is necessary and sufficient 
to show that the variation of the constraint 
$\Ups = 1 + \Phi - ({*}F_3)^2 \approx 0$ is zero. However, we do not
know the form of the function $\Phi$. The way out of this impasse
\cite{Cederwall:1998b,Cederwall:1998a} is to note that if we assume that the 
scalar functional $\Phi$ is formed out of contractions between the
world-volume field strengths and the metric only,\footnote{We thus 
exclude WZ-type terms linear in $\ep^{ijk}$, which is reasonable since in the
formulation we are using such terms are contained in $F_3$.}
a simple scaling argument shows that the variation of $\Ups$ can be 
written\footnote{The part of this expression that is proportional to the 
dilaton variation was derived by temporarily reinstating the suppressed
dilaton-dependence of the action~(\ref{D2ansatz}) by means of the previously
given substitution rules
$F_1\rar e^{\frac{3}{4}\phi}F_1$, $F_2\rar e^{-\frac{1}{2}\phi}F_2$ and 
$F_3\rar e^{\frac{1}{4}\phi}F_3$.}
\bea
\de_{\ka} \Ups &=& K^{i}\,\de_{\ka}F_{i} +
\mbox{$\frac{1}{2!}$}K^{ij}\,\de_{\ka}F_{ij} +
\mbox{$\frac{2}{3!}$}F^{ijk}\,\de_{\ka}F_{ijk} - \big[\half K^{(i}F^{j)} +
\mbox{$\frac{1}{2!}$}{K^{(i}}_{l}F^{j)l} \non\\ && +
\mbox{$\frac{3}{2}$}\scal \mbox{$\frac{2}{3!}$}{F^{(i}}_{lm}F^{j)lm} 
\big]\,\de_{\ka}g_{ij} +
\big[ \mbox{$\frac{3}{4}$}K_1\scal F_{1} -\half K_2\scal F_{2} 
+ \mbox{$\frac{2}{4}$} F_3\scal F_3\big]\,\de_{\ka}\phi\,.
\eea
This improves matters considerably since from~(\ref{D2dualrel}) we have 
explicit expressions for $K_1$ and $K_2$, which are valid when the duality 
relations are imposed. Hence, we are in the fortunate situation that
although we do not know the action we know its variation under a 
$\ka$-transformation. By demanding that 
$\Ups$ be $\ka$-invariant, we can exploit this knowledge to derive the action,
as we shall demonstrate next.

Inserting the expressions for the variations of the world-volume fields,
together with the background constraints leads to
\bea
(\de_\ka\Ups)^{(1/2)} &=& -2\,{*}F_3\,\bLa\Big[ \Xi + 
2\,{*}F_1\scal\ga_2\ga_{11} +
3\,{*}F_2\scal\ga_1 \ga_{11} + (3{-}x)\,{*}(F_1\we F_2) + {*}F_3 \Big]
\ka\,, \non\\
(\de_\ka\Ups)^{(0)} &=& 4i\,{*}F_3\,\bE_i
\Big[\frac{\ep^{ijk}}{2\sqrt{-g}}\ga_{jk}
- {*}F_1^{ij}\,\ga_j\ga_{11} + {*}F_2^{i}\,\ga_{11} +
x\,g^{ij}\,F_1\scal{*}F_2\,\ga_j + F_1^{(i}{*}F_2^{j)}\,\ga_j \non \\ 
&&\hspace{1.2cm} +\,\, g^{ij}\,{*}F_3\,\ga_{j}\Big] \ka \,.
\eea
These expressions should vanish when $\ka$ is of the form $P_+\ze = \half(\id +
\Ga)\ze$, which leads us to another question: how do we determine $P_+$? 
This is a less serious problem since there are not that many ``natural'' 
terms that 
can appear. In all known cases one can write ${*}P_+$ as a $p{+}1$-form
expression involving the world-volume fields and the totally
antisymmetric products of $\ga$-matrices, $\ga_{i_1\ldots
 i_r}$, considered as forms. By making a natural Ansatz for $P_+$,
inserting it into the above variations, and then using the 
formula~(\ref{general}) to expand the products of $\ga_{i_1\ldots i_r}$'s
one obtains two expressions which are linear combinations of
$\ga_{i_1\ldots
 i_r}$'s, for various values of $r$. Requiring these expressions to vanish
forces each tensor component to vanish separately, since the $\ga_{i_1\ldots
 i_r}$'s are linearly independent (for generic embeddings). 
These conditions determine expressions for the duality 
relations~(\ref{D2dualrel}), which can be integrated to give $\Phi$; 
for further details, see appendix~\ref{kappadetails}. 
The projection operator which makes the above variation vanish is 
\be
2\,{*}F_3\,P_\pm = {*}F_3\,\id 
\mp\big[\Xi - x\,{*}F_1\scal \ga_2\ga_{11} +
(1{-}x)\,{*}F_2\scal\ga_1 \ga_{11} \big]\,,
\ee
and the result for the action is
\be
S = \int \D^3\xi\,\sqrt{-g}\,\la\,\big[1 + x\,F_1\scal F_1 + 
(1{-}x)\,F_2\scal F_2 +
x\,(1{-}x)\,F_1\scal F_1\;F_2\scal F_2 - ({*}F_3)^2\big]\,, 
\ee
supplemented by the duality relations 
\bea
{*}F_3\,{*}F_1 &=& F_2 + x\,(F_1\scal F_1)\,F_2 \,,\non \\
-{*}F_3\,{*}F_2 &=& F_1 + (1{-}x)\,(F_2\scal F_2)\,F_1 \,.
\eea 
At this point we would like to make a few comments. Notice the
similarity between the Bianchi identity for $F_3$ and the
form of ${*}P_+$, which becomes even closer if we use the
formal rules $R_{2n}
\rar -(-)^{n}\ga_{2n-1}(\ga_{11})^{n}$, $H_3 \rar \ga_2\ga_{11}$. 
A similar correspondence holds for all hitherto known
brane actions. If assumed to hold in all cases, it reduces the amount of 
guesswork involved (basically we only had to guess the form for $P_+$)
and makes the method more algorithmic. 
We will comment further on this issue in later sections. 
Let us also remark that in the simple case considered above one could
alternatively have made progress by making an Ansatz for $\Phi$; 
for higher-dimensional cases this approach is less feasible, however. 

To summarise, in the two limits $x=0$ and $x=1$ we recover known results with 
correct expressions for the projection operator $P_+$. For other values of $x$
 we obtain new $\ka$-symmetric formulations of the D2-brane action. 
In particular, for the choice $x=\half$ one obtains a formulation which treats
the two world-volume fields in a symmetric fashion.
We would like to stress that these actions are all equivalent 
(i.e., they describe the same physical object) as is
obvious from the way the parameter $x$ was introduced.

\setcounter{equation}{0}
\section{Some further examples}
\label{IIBbranesect}

\subsection*{The type {\II}B D-string}

The D-string in the type {\II}B theory couples minimally to the two-form 
potential $C_2$. Possible world-volume fields satisfying Bianchi
identities 
are 
\bea
F_0 &=& \mu - C_0\,, \phantom{\D A_1 - B_2} \D F_0 = R_1\,, \non \\
F_2 &=& \D A_1 - B_2\,, \phantom{\mu - C_0}\D F_2 = H_3\,,
\eea
where $\mu$ is a constant. In order for $F_0$ to be invariant under the
Peccei-Quinn symmetry, $\mu$ has to change to compensate for the constant shift
of $C_0$. We will comment on this fact later on. Given the above two fields,
one can construct the following expressions for the tension form $\tF_2$:
\bea
\tF_2 &=& \D \tA_1 - C_2 + x\,B_2\, F_0 - (1{-}x)\,C_0 \,F_2\,, \non \\
\D \tF_2 &=& -R_3 + H_3\,F_0 - (1{-}x)\,R_1\, F_2 \,.
\eea
Here the background field strength $R_3$ is defined as 
\be
R_3 = \D C_2 + x\,B_2 \,\D C_0 - (1{-}x)\,C_0 \,\D B_2\,,
\ee
and satisfies the usual Bianchi identity, $\D R_3 = R_1\, H_3$. 
The action is taken to be of the form\footnote{As for the 
D2-brane we suppress the dilaton dependence; it can be reinstated by using 
the rules: $F_0\rar e^{-\phi}F_0$, $F_2 \rar e^{-\frac{1}{2}\phi}F_2$ and 
$\tF_2 \rar e^{\frac{1}{2}\phi}\tF_2$.}
\be
S = \int \D ^2 \xi \,\sqrt{-g}\, \la \,\big[ 1 + \Phi(F_0, F_2) - 
({*}\tF_2)^2\big]\,.
\ee
In order to be able to apply the method of the previous section it is 
convenient to formally regard $\mu$ as the exterior derivative of a 
``$-1$''-form, $A_{-1}$. This procedure will be justified later in this
section. Using the just mentioned formal trick, the duality relations can 
be shown to be determined by 
\bea
2(1{-}x)\,{*}\tF_2\,{*}F_0 &=& K_2 := \frac{\de \Phi}{\de F_2}\,,\non \\
 -2x\,{*}\tF_2\,{*}F_2 &=& K_0 := \frac{\de \Phi}{\de F_0}\,.
\eea
As in the case of the D2-brane, the $\ka$-variation of the constraint 
$\Ups = 1+\Phi-({*}\tF_2{*})^2\approx 0$ can be written in terms of the $K$'s:
\bea
\de_{\ka} \Ups &=& K_0\,\de_{\ka}F_0 + \mbox{$\frac{1}{2!}$}\,K^{ij}\,
\de_{\ka}F_{ij} + \mbox{$\frac{2}{2!}$}\,\tF^{ij}\,\de_{\ka}\tF_{ij} -
\big(\half\,{K^{(i}}_{l}\,F^{j)l} + {\tF}{}^{(i}{}_{l}\,\tF^{j)l}\big)\,
\de_{\ka}g_{ij} \non\\ && + \mbox{$\frac{1}{2!}$}\,\big( 
\half\tK^{ij}\,\tF_{ij} -\half K^{ij}\,F_{ij}\big)\,\de_{\ka}\phi\,.
\eea
The $\ka$-variations of the fields are $\de_{\ka}F_0 = -i_{\ka}R_1$,
$\de_{\ka}F_2 = -i_{\ka}H_3$ and $\de_{\ka}\tF_2 = -i_{\ka}R_3 +
F_0\,i_{\ka}H_3 - (1{-}x)\,F_2\,i_{\ka}R_1$ (the variations of the metric and
the dilaton are the same as in the D2-brane case). Using these
expressions together with the supergravity constraints (see 
appendix~\ref{sugraapp}) one then obtains the expressions
\bea
(\de_{\ka}\Ups)^{(1/2)} &=& 2\,{*}\tF_2\,\bLa \Big[-\Xi\,J + F_0\,\Xi\,K 
 - \,{*}F_2\,I - (1{+}x)\,F_0\,{*}F_2 - {*}\tF_2 \Big]\ka\,, \non\\
 (\de_{\ka}\Ups)^{(0)} &=& 4i\,{*}\tF_2\,\bE_i \Big[
\frac{\ep^{ij}}{\sqrt{-g} }\,\ga_{j}\,J + F_0\,\frac{\ep^{ij}}{\sqrt{-g}} 
\ga_{j}\,K - (1{-}x)\,F_0\,{*}F_2\,g^{ij}\,\ga_j \non \\ 
&&\hspace{1.7cm}+\,\, g^{ij}\,{*}\tF_2\,\ga_j \Big]\ka\,.
\eea
The projection $\ka=P_+\ze$ which makes the above variations vanish turns out 
to be
\be
P_\pm \propto -{*}(\tF_2 - \al F_0\,F_2)\id \pm [\Xi\,J + (x{+}\al)\,F_0\, 
\Xi\,K + (1{-}(x{+}\al))\,{*}F_2\,I]\,,
\ee
where the parameter $\al$ is undetermined by the requirement of 
$\ka$-symmetry.  
Following the same approach as for the D2-brane one obtains the 
action\footnote{Note the similarity with the result for the D2-brane in the 
previous section, a fact which should be derivable from T-duality 
considerations.}
\be
S = \int \D^2 \xi\,\sqrt{-g}\,\la\,[1 + x\,F_0^2 + (1{-}x)\,F_2\scal F_2 
+ x\,(1{-}x)\,F_0^2\,F_2\scal F_2]\,.
\ee
The duality relations which supplement this action are
\bea
{*}\tF_2\,{*}F_0 &=& F_2 + x\,F_0^2\,F_2\,, \non \\
-{*}\tF_2\,{*}F_2 &=& F_0 + (1{-}x)\,F_2\scal F_2\,F_0 \,.
\eea
It is fairly straightforward to show that the second relation follows
from the first one and the constraint $1 + \Phi = ({*}\tF_2)^2$, 
thus showing that it can be consistently imposed and justifying the earlier 
formal derivation. 

Let us close this subsection by discussing some simple limiting cases of the
above action. When $x=0$, we recover the conventional $\ka$-symmetric D-string
action. In the opposite limit, $x=1$, $\Phi$ equals 
$1 + F_0^2 = 1 + (\mu - C_0)^2$. This result agrees with the well-known result
of refs~\cite{Tseytlin:1996,Aganagic:1997,Oda:1998}, 
where it was shown that $\mu$ has 
the correct transformation property under the Peccei-Quinn symmetry (which in
particular transforms $C_0$ to $C_0 + 1$) to make $F_0$ invariant. It was
furthermore shown that the action could be interpreted as the action for the
fundamental string in an SL(2,$\Z$)-transformed background, thus
establishing the S-duality connection between the D- and F-strings at the
level of their world-volume actions.

\subsection*{The type {\II}B D3-brane}

The complete $\ka$-symmetric action for the super-D3-brane in the type
{\II}B theory was constructed in ref.~\cite{Cederwall:1996b}. It is known 
that the D3-brane is a singlet under the SL(2,$\Z$) symmetry of the type {\II}B
theory. This fact was demonstrated at the level of the world-volume action in
refs~\cite{Tseytlin:1996,Green:1996a,Aganagic:1997,Kimura:1999a}, and 
an action which makes the SL(2,$\Z$) symmetry manifest was constructed in 
ref.~\cite{Cederwall:1998a}. 

In this section we will apply our method to the type {\II}B D3-brane, and
compare our findings with previously known results. The parameterisations for 
the background field strengths which we will use are
\bea
R_1 &=& \D C_0\,, \qquad\qquad\hspace{.5pt} 
R_3 = \D C_2 - C_0\, H_3\,, \non \\ 
H_3 &=& \D B_2\,, \qquad\qquad
R_5 = \D C_4 + x\,B_2 \,\D C_2 - (1{-}x)\,C_2\, H_3 \,.
\eea
Notice that we have not introduced a parameter in the definition for
$R_3$, in contrast to the case of the D-string. We only
introduce parameters which lead to (Poincar\'e) dual
pairs of world-volume field strengths. Although this is {\it a priori}\/  a
restriction, it is unclear whether it excludes any physically interesting 
cases. The introduction of a parameter in the expression for $R_3$ would
have lead to the appearance of $F_0$ in various places, but there is
no corresponding four-form to which it can
be dual (recall that for the D-string there where two
two-forms). Given the above parameterisation the possible gauge
invariant world-volume fields and their Bianchi identities become 
\bea
\label{D3WVfields}
F_{2} &=& \D A_1 - B_2\,,\phantom{\D\tA_1-C_2-C_0\,F_2\,,} \D F_2 = -H_3\,,
 \non\\
\tF_{2} &=& \D \tA_1 - C_2 - C_0\, F_2\,,\phantom{\D A_1-B_2\,,}\D
 \tF_2 = -R_3 -R_1\,F_2\,, \non \\
F_{4} &=& \D A_3 - C_4 +x\,B_2\,\tF_2 - (1{-}x)\,C_2\,F_2 + x\,C_0\,B_2\,F_2 +
(x{-}\half)\,C_0\,F_2\,F_2\,, \non\\
\D F_4 &=& -R_5 + x\,H_3\,\tF_2 - (1{-}x)\,F_2\,R_3 + 
(x{-}\half)\,R_1\,F_2\,F_2\,.
\eea
There is an alternative way of looking at the parameter $x$ in $F_4$
above. Consider the expression $F^{\rm D3}_4 - x\, F_2\,\tF_2$, where 
$F^{\rm D3}_4$ 
is the canonical D-brane expression, $F^{\rm D3}_4 = \D A_3 - C_4 - 
C_2\,F_2$. By making the
field redefinitions $C_4 \rar C_4 + x\,B_2\,C_2$ together with 
$A_3 \rar A_3 - xA_1\,d\tA_1$, we recover the expression for $F_4$ 
given above. Thus, at the 
world-volume level the dualisation is performed by adding the term
$F_2\,\tF_2$ to $F_4$. This is analogous to the way one usually
proceeds. Looking at the dualisation this way leads to an easy
way of determining which field parameterisations are needed for the
background fields.

The variations of the world-volume form fields under a $\ka$-symmetry
transformation are
\bea
\label{D3WVvar}
\de_{\ka}F_2 &=& -i_{\ka}H_3\,, \non \\ 
\de_{\ka}\tF_2 &=& -i_{\ka}R_3 - F_2\,i_{\ka}R_1\,, \non\\
\de_{\ka}F_4 &=& -i_{\ka}R_5 + x\,\tF_2\,i_{\ka}H_3 - (1{-}x)\,F_2\,i_{\ka}R_3
- (\half{-}x)\,F_2\,F_2\,i_{\ka}R_1\,.
\eea 
We make a general Ansatz for the action of the form\footnote{The dilaton
dependence is as usual suppressed, but can be reinstated using 
the rules: $F_2 \rar e^{-\frac{1}{2}\phi}F_2$, 
$\tF_2 \rar e^{\frac{1}{2}\phi}\tF_2$ and $F_4\rar F_4$.}
\be
\label{D3ansatz}
S = \int \D^4 \xi\, \sqrt{-g}\,\la \left[ 1 + \Phi(F_2,\tF_2) - 
({*}F_4)^2\right]\,.
\ee
It is often convenient to rewrite this action in ``form language'' as
\be
S = - \int \la \left[ {*}1 + {*}\Phi(F_2,\tF_2) + F_4\,{*}F_4\right]\,.
\ee
This form of the action is better suited for the derivation of the duality
relations consistent with the Bianchi identities and the equations of motion. 
In the present case these duality relations must be of the form
\bea
\label{D3dualrel}
2(1{-}x)\,{*}F_4\,{*}\tF_2 &=& K_2:=\frac{\de \Phi}{\de F_2}\,,\non\\
 -2x\,{*}F_4\,{*}F_2 &=& \tK_2 := \frac{\de \Phi}{\de \tF_2}\,.
\eea
In order for the action~(\ref{D3ansatz}) to be $\ka$-invariant
it is necessary and sufficient that the constraint $\Ups = 1 +
\Phi(F_2,\tF_2) - ({*}F_4)^2 \approx 0$ is invariant.
By using the same argument as before we can rewrite the
$\ka$-variation of the constraint in terms of the $K$'s as 
\bea
\de_{\ka} \Ups &=& \mbox{$\frac{1}{2!}$}K^{ij}\de_{\ka}F_{ij} +
\mbox{$\frac{1}{2!}$}\tK^{ij}\de_{\ka}\tF_{ij} +
\mbox{$\frac{2}{4!}$}F^{ijkl}\de_{\ka}F_{ijkl} - \big(\half
{K^{(i}}_{l}F^{j)l} + \half{\tK^{(i}}{}_{l}\tF^{j)l} \non\\ && +\, 
\mbox{$\frac{1}{3!}$}{F^{(i}}_{lmn}F^{j)lmn} \big)\de_{\ka}g_{ij} +
\mbox{$\frac{1}{2!}$}\big( \half\tK^{ij}\tF_{ij} -
\half K^{ij}F_{ij}\big)\de_{\ka}\phi\,.
\eea
Inserting the expressions (\ref{D3dualrel}) for the $K$'s and using the 
variations (\ref{D3WVvar}) of the world-volume fields together with the
on-shell constraints for the background fields, we then obtain 
\bea
\label{UpsD3}
(\de_{\ka}\Ups)^{(1/2)} &=& 2\,{*}F_4\,\bLa \Big[
\half\, e^{\phi/2}\,{*}\tF^{ij}\ga_{ij}\,K -
\half\, e^{-\phi/2}\,{*}F^{ij}\ga_{ij}\,J - e^{-\phi}\,{*}(\tF \we \tF)\,I
- {*}(F \we \tF) \Big]\ka\,, \non\\
 (\de_{\ka}\Ups)^{(0)} &=& 4i\,{*}F_4\,\bE_i \Big[
\frac{\ep^{ijkl}}{3!\sqrt{-g} } \ga_{jkl}\,I + e^{\phi/2}{*} \tF^{ij}
\ga_{j}\,K + e^{-\phi/2}\, {*}F^{ij}\ga_{j}\,J \non\\
&&\hspace{1cm}-\,\, (1{-}x)g^{ij}\,{*}(\tF_2\we F_2)\ga_{j} +
{*}{F^{(i}}_{l}\tF^{j)l}\ga_j + g^{ij}\,{*}F_4\ga_j \Big]\ka\,.
\eea
The correct projection operator can be shown to be (more details can be found
in appendix~\ref{kappadetails})
\bea
P_\pm &\propto& {*}\big[F_4 - \al\,(F_2\we \tF_2)\big]\,\id \mp 
\bigg[\Xi I + \half(1 - (x+ \al))\,{*}F_2^{ij}
\ga_{ij}J \,,
 \non \\ &&\hspace{1cm}+\,\,\half(x{+}\al)\,{*}\tF_2^{ij}\ga_{ij}K + 
\half(1-2(x{+}\al))\,{*}(F_2\we F_2)I \bigg]\,.
\eea
Again we note the similarity with the Bianchi identity; from this 
perspective the free 
parameter $\al$ can be understood from the fact that the Bianchi
identity for $F_4$ can be rewritten as \,
$\D (F_4 - \al F_2\,\tF_2) = -R_5 + (x+\al)H_3\,\tF_2 - [1-(x+\al)]F_2\,R_3 +
[(x+\al)-\half]R_1\,F_2\,F_2\,$. 
Inserting the above expression for $\ka = P_+ \ze$ into the variations 
(\ref{UpsD3}) and using the expansion
formula (\ref{general}) for the product of two $\ga$-matrices, we obtain a 
set of component expressions which must each vanish in order for 
$\ka$-symmetry to hold. The final results of the analysis of these 
expressions are the duality relations (for further details, see appendix
\ref{kappadetails})
\bea
-{*}F_4\,{*}F_2 &=& \tF_2 - (1{-}x)\,{*}(F_2 \we \tF_2)\,{*}F_2
+ \half\,(1{-}x)\,{*}(F_2\we F_2)\,{*}\tF_2 - \half\,x\,{*}(\tF_2\we \tF_2)\,
{*}\tF_2\,, \non\\
{*}F_4\,{*}\tF_2 &=& F_2 - x\,{*}(F_2 \we \tF_2)\,{*}\tF_2 + 
\half\,x\,{*}(\tF_2\we \tF_2)\,{*}F_2-\half(1{-}x)\,{*}(F_2\we F_2)\,{*}F_2\,,
\eea
which can be derived from the $\ka$-invariant action
\bea
S &=& \int \D^4\xi\,\sqrt{-g}\,\la\bigg[1 + (1{-}x)\,F_2\scal F_2 + 
x\,\tF_2\scal\tF_2
-x\,(1{-}x)\,{*}(F_2\we \tF_2)\,{*}(F_2\we\tF_2) \non \\ 
&&\hspace{2cm} +\,\, \half\,x\,(1{-}x)\,{*}(F_2\we
F_2)\,{*}(\tF_2\we\tF_2) -\fourth\,(1{-}x)^{2}\,{*}(F_2\we
F_2)\,{*}(F_2\we F_2) \non \\ &&\hspace{2cm}-\,\,\fourth\,x^{2}\,
{*}(\tF_2\we \tF_2)\,{*}(\tF_2\we \tF_2) \bigg]\,.
\label{D3xaction}
\eea
The value $x=0$ corresponds to the usual D3-brane action with the standard
projection operator, a result we have arrived at using only the requirements 
of $\ka$ and gauge invariance. In this limit there is no need for
the duality relations since they just define $\tF_2$ in terms $F_2$, but 
$\tF_2$ appears neither in the action nor in the projection operator. 
As a consequence, the duality relations can simply be dropped in this case. 
Another interesting limit is $x=1$, in which the action is also of Born-Infeld 
form, but with $\tF_2$ as the world-volume field. However, unlike $\tF_2$
in the limit $x=0$, $F_2$ does
not decouple completely, since the action depends implicitly on $F_2$ through
$\tF_2$ and $F_4$. The explicit dependence of $F_2$ in 
$P_+$ can be removed by using the identity (\ref{D34formid}), but the implicit
dependence remains. It is nevertheless true
that the equation of motion for $A_1$ becomes non-dynamical, as can
be demonstrated in the following way.
When $x=1$, the equation of motion for $A_1$ is 
\be
\D \left[ \la\left\{ 2{*}F_4\,F_2- \frac{\de {*} \Phi}{\de \tF_2} \right\}
C_0\right] = 0 \,.
\ee
Here the expression in curly brackets vanishes trivially as a consequence of
the duality relations. 
The action in the limit $x=1$ differs from the dual action derived in 
refs~\cite{Tseytlin:1996,Aganagic:1997}. 
The reason for this discrepancy is that 
the dual field of refs~\cite{Tseytlin:1996,Aganagic:1997} does not satisfy a 
Bianchi identity, nor is it invariant under the Peccei-Quinn symmetry. It thus
violates two of our basic assumptions. We can reproduce the results of 
refs~\cite{Tseytlin:1996,Aganagic:1997} by solving the duality relations to 
eliminate $F_2$ at the level of the equations of motion and then
integrate  
the result to an action; this action agrees with the one derived
earlier using conventional Poincar\'e dualisation.

Yet another interesting case is $x=\half$. This is the most symmetrical choice
for the parameter $x$, and the corresponding action is, as one might suspect, 
related to the manifestly SL(2,$\Z$)-covariant action constructed in
ref.~\cite{Cederwall:1998a}. More precisely, the action (\ref{D3xaction}) with 
$x=\half$ is a gauge-fixed version of the manifestly SL(2,$\Z$)-covariant 
action constructed in ref.~\cite{Cederwall:1998a} and can easily be lifted to 
the latter. In the manifestly covariant formulation the two world-volume field 
strengths are combined into a single complex field, 
${\mathcal F}= {\mathcal U}^r F_{2;r} $, which by construction is invariant 
under SL(2,$\Z$) but transforms under the (local) U(1) symmetry of the 
background theory.\footnote{See appendix \ref{sugraapp} for a brief
account of some relevant properties of type {\II}B supergravity.}
Here $F_{2;r}=\D A_{1;r}-C_{2;r}$ transforms as a doublet under 
SL(2,$\Z$), where $C_{2;1}=B_2$, $C_{2;2}=C_2$
and similarly for $A_{1;r}$. By choosing the U(1) gauge 
${\mathcal U}^{1}=-e^{\frac{1}{2}\phi} C_0+ie^{-\frac{1}{2}\phi}$
and ${\mathcal U}^2=e^{\frac{1}{2}\phi}$, one gets 
${\mathcal F} = e^{\frac{1}{2}\phi}\tF_2 + i e^{-\frac{1}{2}\phi}F_2$. 
Implementing this relation in the action leads to complete agreement with 
the results of ref.~\cite{Cederwall:1998a} (after taking into account the 
differences in conventions for $F_4$ and $R_5$). We have thus explicitly 
shown in a simple way that the usual D3-brane action and the manifestly 
covariant action of ref.~\cite{Cederwall:1998a} describe the same physical 
object---the D3-brane---which is a singlet under SL(2,$\Z$).

\setcounter{equation}{0}
\section{The type {\II}B 5-branes}
\label{IIB5branesect}

In type {\II}B supergravity in its doubled formulation there are two six-form 
gauge potentials, $C_6$ and $B_6$, which are dual to the two-form potentials
$C_2$ and $B_2$, respectively. The six-form potentials couple minimally 
to two different branes: the D5-brane and the NS5-brane. Below we
discuss these two objects using the method outlined in previous sections.
We will be somewhat more elaborate in our treatment of the NS5-brane
as this is the most interesting and novel case. 

\subsection*{The D5-brane}
\label{IIBD5sect}

The D5-brane couples minimally to the six-form potential $C_6$. The
standard form of the associated world-volume six-form is
\cite{Bergshoeff:1998c} 
\be
F_6 = \D A_5 - C_6 -C_4\,F_2 -
\half\,C_2\,F_2\,F_2 - \sixth\,C_0\,F_2\,F_2\,F_2\,.
\ee
In addition to this tension form only $F_2$ appears in the action. 
As in previous sections, this action can be
generalised by introducing parameters into the definitions of the
background field strengths. In analogy with the case of the D3-brane, 
we only introduce dual pairs of world-volume fields, i.e.,
fields that are related by auxiliary duality relations. In other words,
we are dualising the two-form field
strength $F_2$ into the four-form field strength $F_4$. 
In world-volume language this means that we are redefining $F_6$ as 
$F_6 \rar F_6 - y\,F_2\,F_4$. 
We let the background fields $R_1$, $R_3$ and $R_5$ have the canonical
form $R = \D C - C\, H$, whereas $R_7$ is given by
\bea
R_7 &=& \D C_6 + y\, \D C_4\, B_2 -(1{-}y)\, C_4\,H_3\,.
\eea
These field strengths all obey Bianchi identities of the form $\D R-R\,H=0$.
The corresponding world-volume fields are
\bea
F_{2} &=& \D A_{1} - B_{2}\,, \non\\
F_{4} &=& \D A_{3} - C_{4} - C_{2}\,F_{2}
-\half \,C_{0}\,F_{2}\,F_{2}\,, \non\\
F_{6} &=& \D A_{5} - C_{6} + y\,B_{2}\,F_{4} -
(1{-}y)\,C_{4}\,F_{2} - (\half{-}y)\, C_{2}\,F_{2}\,F_{2}\,, \non\\ && +
y\, B_{2}\,C_{2}\,F_{2} - \half (\third {-}y)\,
C_{0}\,F_{2}\,F_{2}\,F_{2}
+\half\,y\,C_{0}\,B_{2}\,F_{2}\,F_{2} \,,
\eea
satisfying the Bianchi identities
\bea
\D F_{2} &=& - H_{3}\,, \non\\
\D F_{4} &=& -R_{5} - R_{3}\,F_{2} -\half\, R_{1}\,F_{2}\,F_{2}\,,\non \\
\D F_{6} &=& -R_7 + y\,H_3\,F_4 - (1{-}y)\,R_5\,F_2 - 
(\half{-}y)\,R_3\,F_2\,F_2 - 
\half\,(\third {-}y)\,R_1\,F_2\,F_2\,F_2\,.
\eea
To verify that $F_4$ is an appropriate dual field in the conventional sense
one can apply the methods of refs~\cite{Tseytlin:1996,Aganagic:1997} to the 
usual DBI D-brane action. One then
obtains exactly the first duality relation given in (\ref{D5dualrel}) below,
after the
identification ${*}F_6 = \sqrt{1+\Phi(F_2)}$, where 
$\Phi(F_2)=\frac{\det(g+F)}{\det(g)}-1$.

Our Ansatz for the action is
\be
S = \int \la\,[{*}1 + {*}\Phi(F_{2},F_{4}) + F_6\,{*}F_6]\,.
\ee
The duality relations compatible with the Bianchi identities and the
equations of motion can then be shown to be 
\bea
\label{D5dualrel}
 -2y\,{*}F_{6}\,{*}F_{2} &=& K_{4} := \frac{\de\Phi}{\de F_4}\,,\non\\
 2(1{-}y)\,{*}F_{6}\,{*}F_{4} &=& K_{2} := \frac{\de\Phi}{\de F_2}\,.
\eea
Furthermore, the variation of the constraint $\Ups = 1 + \Phi(F_{2},F_{4})-
({*}F_{6})^2\approx0$ is\footnote{As usual we make the assumption that $\Phi$
can be constructed from only the form fields and the metric. Moreover, the 
dilaton variation is obtained using the rules 
$F_2\rar e^{-\frac{1}{2}\phi}F_2$, $F_4\rar e^{0{\cdot}\phi}F_4$ and 
$F_6\rar e^{-\frac{1}{2}\phi}F_6$.} 
\bea
\de_{\ka}\Ups &=& \mbox{$\frac{1}{2!}$}\,K^{ij}\,\de_{\ka}F_{ij} +
\mbox{$\frac{1}{4!}$}\,K^{ijkl}\,\de_{\ka}F_{ijkl} +
\mbox{$\frac{2}{6!}$}\,F^{ijklmn}\,\de_{\ka}F_{ijklmn} -
\half\,{K^{(i}}_{l}\,F^{j)l}\,\de_{\ka}g_{ij} \non\\ 
&&- \mbox{$\frac{1}{12}$}\,{K^{(i}}_{lmn}\,F^{j)lmn}\,\de_{\ka}g_{ij} -
\mbox{$\frac{1}{5!}$}\,{F^{(i}}_{lmnpq}\,F^{j)lmnpq}\,\de_{\ka}g_{ij} -
\fourth\,K^{ij}\,F_{ij}\,\de_{\ka}\phi \non\\
&&- \mbox{$\frac{1}{6!}$}\,F^{ijklmn}\,F_{ijklmn}\,\de_{\ka}\phi\,.
\eea

By inserting the explicit expressions (\ref{D5dualrel}) for the $K$'s, 
as well as the supergravity on-shell constraints (see appendix 
\ref{sugraapp}), we obtain 
\bea
\label{D5variations}
(\de_{\ka}\Ups)^{(1/2)} &=& 2\,{*}F_6\,\bLa \Big[\Xi\,J + {*}F_4\scal \ga_2\,K
 - \half\,{*}(F_2\we F_2)\scal \ga_2\,J - \third\,{*}(F_2\we F_2
 \we F_2)\,I \non\\
&&- (1{-}y)\,{*}(F_2\we F_4) + {*}F_6 \Big]\ka\,, \non\\
(\de_{\ka}\Ups)^{(0)} &=& 4i\,{*}F_6\,\bE_i \Big[({*}\ga_5)^i\,I +
 \mbox{$\frac{1}{3!}$}\,{*}F_2^{ijkl} \ga_{jkl}\,I + {*}F_4^{ij}\ga_{j}\,K +
 \half\,{*}(F_2\we F_2)^{ij}\ga_j\,J \non\\
&&+ y\, g^{ij}\,{*}(F_2\we F_4)\ga_{j} + {*}{F_4^{(i}}_{l}F_2^{j)l}\ga_j +
g^{ij}\,{*}F_6\,\ga_j \Big]\ka\,.
\eea
In analogy with the previously considered cases, the calculation proceeds by 
inserting the projected spinor-parameter $\ka=P_+\ze$ using an appropriate
Ansatz for the projection operator and examining the irreducible components
of the expression obtained by application of the $\ga$-matrix product identity
(\ref{general}); for details we refer to appendix \ref{kappadetails} quoting 
here only the results.
 
The projection operator is found to be
\bea
\label{D5projop}
P_\pm &\propto& - {*}(F_6- \al\,F_2\we F_4)\,\id \pm \big[\, \Xi\,J + 
(1{-}(y{+}\al))\,{*}F_2\scal \ga_4\,I +
(y{+}\al)\,{*}F_4\scal \ga_2\,K \non\\ &&
+\,\, (\half{-}(y{+}\al))\,{*}(F_2\we F_2)\scal \ga_2\,J 
 + \half\,(\third {-}(y{+}\al))\,{*}(F_2\we F_2 \we F_2)\,I\,\big]
\,.
\eea
Here we note, in particular, that the correspondence with the tension field 
strength holds also in this case. Moreover, the explanation for the free 
parameter $\al$ is the same as in the D3-brane case. 

The final expression for the action is
\bea
\label{D5action}
S &=& \int\D^6\xi\,\sqrt{-g}\,\la\,\bigg[ 1 + (1{-}y)\,F_2\scal F_2 + 
y\,F_4\scal F_4 - y\,(1{-}y)\,{*}(F_2 \we F_4)\,{*}(F_2\we F_4) \non\\
&& \qquad + \half\, y\,(F_2\we
F_2)\scal({*}F_4\we \,{*}F_4) + \half(\half{-}y)\,(F_2 \we F_2)\scal
(F_2\we F_2) \non\\
&& \qquad - \mbox{$\frac{1}{12}$}(\third {-}y)\,{*}(F_2\we F_2 \we
F_2)\,{*}(F_2\we F_2\we F_2)- ({*}F_6)^2\bigg]\,,
\eea
which is to be supplemented with the duality relations
\bea
\label{D5dualrels}
-{*}F_6\,{*}F_2 &=& F_4 - (1{-}y)\,{*}(F_2\we F_4)\,{*}F_2 + \half\,{*}(F_2\we
F_2)\we\,{*}F_4\,, \non\\
{*}F_6\,{*}F_4 &=& F_2 - y\,{*}(F_2\we F_4)\,{*}F_4 
- \fourth\,{*}[{*}(F_2\we F_2+{*}F_4\we{*}F_4)\we F_2]\non \\
&& -\fourth\mbox{$\frac{1-3y}{1-y}$}\,{*}\bigg\{
 {*}[F_2\we F_2 -{*}F_4\we{*}F_4 - \third\,{*}(F_2\we F_2\we
F_2)\,{*}F_2] \we F_2 \bigg\} \,.
\eea
In the last line of the second relation we have extracted a linear combination
of terms which can be shown to vanish identically when both duality relations 
are satisfied (see appendix~\ref{kappadetails}, 
in particular eq.~(\ref{D5id})). 
Hence, in applications of the D5-brane formulation under discussion
one may simply drop these terms and use the simplified expression.

As expected, we recover the usual D5-brane action for $y=0$. 
In the opposite limit, $y=1$, we get a dual (but equivalent) description. 
However, $F_2$ does not decouple completely from the action in this limit; 
unlike the situation for the D3-brane, in addition to the implicit dependence,
$F_2$ also appears explicitly in the action. 
In order to fully eliminate $F_2$ one needs to solve an algebraic 
equation similar to the one encountered in ref.~\cite{Aganagic:1997}. 
The difference compared to earlier approaches is that we have the additional 
information of knowing the equations of motion and the action of the dual 
theory, so we can in principle determine the dynamics of the
three-form potential. 
Another interesting limit appears to be $y=\third $, where the
action becomes quartic. One would like to relate the action for the
D5-brane given above to an action for the NS5-brane; we will return to
this issue once we have constructed an action for the NS5-brane.

\subsection*{The NS5-brane}

It has been known for several years \cite{Callan:1991} that the world-volume 
field theory of the type {\II}B NS5-brane is described (on-shell) by a
six-dimensional vector multiplet. 
However, there has been a debate in the literature about whether the 
action should be formulated in terms of a two-form field strength 
(as for the D5-brane) or in terms of a four-form field strength (in keeping
with the prescription of accompanying a target-space S-duality transformation
with a world-volume Poincar\'e duality transformation); for discussions see 
e.g.\ refs~\cite{Bergshoeff:1996a,Hull:1997}. It is
furthermore known that the D5-, D3- and D1-branes can end on the NS5-brane
(see e.g.\ ref.~\cite{Argurio:1997}). From this perspective one expects a 
six-form, a four-form and a two-form to be present in the world-volume theory 
of the NS5-brane. The six-form in question is different from the one 
describing the tension and the need for it has been discussed in 
ref.~\cite{Bergshoeff:1998a}. 
Below we will construct an action whose world-volume field content reflects 
the above mentioned facts.

First we need to choose a proper parameterisation of the background
field strengths. Since it is known that the fundamental string cannot
end on the NS5-brane, we use as a guiding principle the condition that
$F_2$ should not appear (not even implicitly) in the world-volume
action. This restriction leads to the following expressions for the RR
background field strengths:
\bea
R_1 &=& \D C_0 \,, \non\\
R_3 &=& \D C_2 + B_2\, \D C_0 \,,\non\\
R_5 &=& \D C_4 + B_2\,\D C_2 +\half\, B_2\, B_2 \,\D C_0\,, \non\\
R_7 &=& \D C_6 + B_2 \,\D C_4 + \half\, B_2\, B_2 \,\D C_2 + 
\sixth\, B_2\, B_2 \, B_2\, \D C_0 \,.
\eea
The corresponding world-volume field strengths become\footnote{Here, 
as in the case of the D-string, 
$\D A_{-1}$ formally denotes an exact zero-form, i.e.\ a constant.} 
\bea
F_0 &=& \D A_{-1} -C_0\,, \non\\ 
\tF_2 &=& \D \tA_1 - C_2 + F_0\, B_2 \,,\non\\
F_4 &=& \D A_3 - C_4 + B_2\,\tF_2 - \half\, B_2\, B_2\, F_0 \,,\non\\
F_6 &=& \D A_5 - C_6 + B_2\, F_4 - \half\, B_2\, B_2\, \tF_2
+ \sixth\,B_2\, B_2\, B_2\, F_0 \,.
\eea
These relations can be succinctly written as $R = e^{B}\,\D C$ and $e^{-B}F =
 \D A - C$,
where the last definition is iterative and $F = F_0 + \tF_2 + F_4 + F_6$. 
In the same compact notation the Bianchi identities are
\be
\D F = - R + H_3\, F\,.
\label{NS5wvBIs}
\ee
The gauge-invariant field strength of the NS-NS six-form gauge potential
is chosen as 
\bea
H_7 &=& \D B_6 - (1{-}y)\,C_6\,\D C_0 + y\,C_0\,\D C_6 - (1{-}x)\,C_2\, \D C_4 
+ x\, C_4\,\D C_2 \,,\non\\
\D H_7 &=& R_7\,R_1 - R_5 \,R_3\,.
\eea
The associated tension form is then found to be
\bea
\tF_6 &=& \D \tA_5 - B_6 - (1{-}y)\,C_6\,F_0 + y\,C_0\,F_6 - (1{-}x)\,C_2\,F_4
 + x\,C_4\,\tF_2 + (1{-}x{-}y)\,B_2\,F_0\,F_4 \non\\
&&+ (\half{-}x)\,B_2\,\tF_2\,\tF_2 +
(1{-}x)\,B_2\,C_2\,\tF_2 - x\,B_2\,C_4\,F_0 - y\,B_2\,C_0\,F_4 \non\\ 
&&-(1{-}\mbox{$\frac{3}{2}$}x{-}\half y)\,B_2\,B_2\,F_0\,\tF_2 - 
\half(1{-}x)\,B_2\,B_2\,C_2\,F_0 + \half\,y\,B_2\,B_2\,C_0\,\tF_2 \non\\
&& + (\third {-}\half x{-}\sixth y)
\,B_2\,B_2\,B_2\,F_0^2 -\sixth\,y\,B_2\,B_2\,B_2\,C_0\,F_0\,,
\eea
with Bianchi identity
\bea
\D \tF_6 &=& -H_7 -(1{-}y)\,R_7\,F_0 + y\,R_1\,F_6 - (1{-}x)\,R_3\,F_4 
+ x\,R_5\,\tF_2
\non\\ &&+ (1{-}x{-}y)\,H_3 \,F_0\, F_4 + (\half{-}x)\,H_3\,\tF_2\,\tF_2\,.
\label{NS5tF6BI}
\eea
The fact that one has a non-dynamical six-form in addition to the
tension form makes the situation analogous to the D-string
discussion, where two field strengths of maximal degree are present
together with a non-dynamical scalar $F_0$.

Given the above field content, the Ansatz for the action takes the by now 
familiar form\footnote{Again, we suppress the dilaton dependence. When 
deriving the $\ka$-variation it is temporarily reinstated by means of
the rules $F_{2k}\rar e^{(1-\frac{1}{2}k)\phi}F_{2k}$ (here $F_2=\tF_2$) and
$\tF_6\rar e^{\frac{1}{2}\phi}\tF_6$.}
\be 
S= \int \la \left[{*}1+{*}\Phi(F_0,\tF_2,F_4,F_6)+\tF_6\,{*}\tF_6\right]\,.
\label{NS5ansatz}
\ee 
Compatibility between the equations of motion and the Bianchi identities
leads to the following expressions for the duality relations:
\bea
\label{NS5dualrels}
2y\,{*}\tF_6\,F_0 &=& {*}K_6 := \frac{\de{*}\Phi}{\de F_6}\,,\non\\
-2(1{-}x)\,{*}\tF_6\,F_2&=&{*}K_4:=\frac{\de{*}\Phi}{\de F_4}\,,\non\\
2x\,{*}\tF_6\,F_4&=&{*}K_2:=\frac{\de{*}\Phi}{\de F_2}\,,\non\\
-2(1{-}y)\,{*}\tF_6\,F_6&=&{*}K_0:=\frac{\de{*}\Phi}{\de F_0}\,.
\eea
The derivation of the last relation is at this point formal, but
will be justified by the final result. 
The total $\ka$-variation of the constraint 
$\Ups = 1 + \Phi(F_0,\tF_2,F_4,F_6) - ({*}\tF_6)^2 \approx 0$ is
\bea
\de_{\ka} \Ups &=& K_0\,\de_\ka F_0 +
\mbox{$\frac{1}{2!}$}\tK^{ij}\,\de_{\ka}\tF_{ij} +
\mbox{$\frac{1}{4!}$}K^{ijkl}\,\de_{\ka}F_{ijkl} +
\mbox{$\frac{1}{6!}$}K^{ijklmn}\,\de_{\ka}F_{ijklmn} +
\mbox{$\frac{2}{6!}$}\tF^{ijklmn}\,\de_{\ka}\tF_{ijklmn} \non\\ &&- \big[
\mbox{$\frac{1}{2!}$}{{\tK}^{(i}}{}_{l}\,\tF^{j)l} +
\mbox{$\frac{2}{4!}$}{K^{(i}}_{lmn}\,F^{j)lmn} +
\mbox{$\frac{3}{6!}$}{K^{(i}}_{lmnpq}\,F^{j)lmnpq} \big]\,\de_{\ka}g_{ij} +
\mbox{$\frac{1}{2!}$}\big[ K_0\,F_0 + \half\,\tK^{ij}\,\tF_{ij} \non\\
&&- \half\,K^{ij}\,F_{ij} -
\half\scal \mbox{$\frac{1}{6!}$}\,K^{ijklmn}\,F_{ijklmn} +
\half\scal \mbox{$\frac{2}{6!}$}\,\tF^{ijklmn}\,\tF_{ijklmn} \big]\,
\de_{\ka}\phi\,.
\eea
Inserting the duality relations (\ref{NS5dualrels}) and the 
expressions for the variations of the world-volume fields with the
background constraints imposed leads to 
\bea
\label{NS5variations}
(\de_{\ka}\Ups)^{(1/2)} &=& 2\,{*}\tF_6\,\bLa \Big[ \Xi\, K + F_0\,\Xi\, J 
- {*}F_4\scal \ga_{2}\,J + [F_0\,{*}F_4-\half{*}(F_2 \we F_2)]
\scal \ga_{2}\,K \non\\ && 
+ 2 \,{*}F_6\,I + (2{-}y)F_0\,{*}F_6 - x \,{*}(\tF_2\we F_4) - 
\,{*}\tF_6 \Big]\ka\,, \non\\
(\de_{\ka}\Ups)^{(0)} &=& 4i \,{*}\tF_6\,\bE_i \Big[
-({*}\ga_5)^i\,K + F_0\,({*}\ga_5)^i\,J
- \mbox{$\frac{1}{3!}$}\,{*}\tF_{2}^{ijkl}\,\ga_{jkl}\,I + 
\,{*}F_{4}^{ij}\,\ga_j\, J \non\\ && + 
[F_0\,{*}F_4^{ij}- \half\,{*}(\tF_2\we \tF_2)^{ij}]\,\ga_j\,K 
+ \big[{-}(1{-}x)\,g^{ij}\,{*}(\tF_2\we F_4) \non\\ && 
+ \,{*}{F_4^{(i}}_l\, \tF_2^{j)l}
+ y\,g^{ij}\,F_0\,{*}F_6+ g^{ij}\,{*}\tF_6\big]\ga_j \Big]\ka\,.
\eea

After a rather long and in parts somewhat intricate calculation (outlined 
in appendix \ref{kappadetails}) it is possible to show that the above 
variations vanish when $\ka=P_+\ze$ and the projection operator is
\bea
P_\pm &\propto& \pm \,\Big[\Xi\,K + (\half{+}3\al)\,F_0\,\Xi\,J + 
(\half{+}\al)\,{*}\tF_2\scal \ga_4\,I - (\half{-}\al) \,{*}F_4\scal \ga_2\,J
\non\\ && -\,\, 2\,\al\, [F_0\,{*}F_4 
-\half\,{*}(\tF_2 \we \tF_2)]\scal \ga_2\,K + (\half{-}3\,\al)\,{*}F_6\,I\Big]
 \non \\
&&+\,\, {*}\big[\tF_6-
(\half{-}3\,\al{-}y)\,F_0\,F_6-(\half{+}\al{-}x)\,\tF_2\we F_4\big]\,\id\,,
\eea
where $\al$ is arbitrary. 
Moreover, the action is given by (\ref{NS5ansatz}) with 
\bea
\label{NS5Phi}
{*}\Phi &=& (1{-}y)\,F_0\,{*}F_0 + x\,\tF_2\we{*}\tF_2 + (1{-}x)\,F_4\we{*}F_4
+y\,F_6\,{*}F_6 \non\\ && - y\,(1{-}y)\,({*}F_6)^2\,F_0\,{*}F_0 
+ x\,(1{-}x)\,(\tF_2\we F_4)\,{*}(\tF_2\we F_4)
\non\\ &&
+ \half\,(1{-}x)\,{*}({*}F_4\we{*}F_4)\we\tF_2\we\tF_2 
+ \half\,(x{-}\half)\,{*}(\tF_2\we\tF_2)\we\tF_2\we\tF_2
\non\\&& +
\third\,y\,{*}F_6\,\tF_2\we\tF_2\we\tF_2  
-\half\,(x{-}y)
\,F_0\,{*}F_4\we\tF_2\we\tF_2 + (1{-}x{-}y)\,(F_0)^2\,F_4\we{*}F_4 \non\\
&&-\sixth\,(2{-}3x{+}y)\,F_0\,{*}({*}F_4\we{*}F_4)\we F_4 -
2\,x\,y\,F_0\,{*}F_6\,\tF_2\we F_4 \non \\ 
&& - \sixth\,(2{-}3x{-}y)\left[{*}(F_0\,F_4 -
\half\tF_2\we\tF_2)\right]^{3} \,.
\eea
Although this expression looks rather intimidating, it 
simplifies somewhat in certain limits of parameter space. 
The duality relations supplementing the action are readily obtained by 
inserting the appropriate functional derivatives of this expression in 
the equations~(\ref{NS5dualrels}); equivalent expressions are
listed in appendix~\ref{kappadetails}. It is also worth noting that
one has the freedom of adding to the action expressions whose variations
vanish as a consequence of the duality constraints, e.g.\ quadratic
combinations of the constraints given in eq.~(\ref{NS5idI}). 

Finally, let us comment on the relation of our formulation of 
the NS5-brane to the one obtained in ref.~\cite{Eyras:1998}, and to
the formulation of the D5-brane above. By inspection
of (\ref{NS5Phi}) one notices that there is no way to consistently eliminate
the four-form field 
strength at the level of the action. In ref.~\cite{Eyras:1998}, T-duality 
considerations led to an action for the NS5-brane which 
is related to the D5-brane by a simple S-duality transformation of
the supergravity background, without the need for any Poincar\'e-duality 
transformations in the world-volume. However, as for the dual D3-brane action
of refs~\cite{Tseytlin:1996,Aganagic:1997}, the world-volume field strength
of ref.~\cite{Eyras:1998} does not satisfy a Bianchi identity, nor is it 
invariant under $C_0 \rar C_0 + 1$; this is the primary 
source for the discrepancy. However, if both actions are to describe the same
object, it must be possible to obtain the results of ref.~\cite{Eyras:1998} 
from ours at the level of the equations of motion. In order to
accomplish this one must eliminate $F_4$ and $F_6$ from the
equation of motion for $\tA_1$. As a first step in this direction one
has to choose $x=1$ and $y=0$. It turns out to be rather involved (if at all 
possible) to algebraically eliminate $F_4$, and we therefore temporarily  
limit our considerations to backgrounds for which $C_0=0$ and 
set $\mu=0$ so that $F_0=0$. Under these simplifying assumptions we
find agreement with the results of ref.~\cite{Eyras:1998}. 
It would be of interest to investigate whether this is still the case when
$C_0\neq0$.
Furthermore, implementing the restriction $F_0=0$ into the expression
for $\Phi$ leads to complete agreement with the action for the D5-brane in 
eq.~(\ref{D5action}), provided we let $x \rar 1-y$, $\tF_2 \rar -F_2$,
$F^{\rm NS}_4 \rar F^{\rm D5}_4$ and $\tF_6 \rar F_6$. 
These transformations follow from making an SL(2,$\Z$) transformation of the 
background; under SL(2,$\Z$) the combinations $(B_2, C_2)$, $(A_1, \tA_1)$, 
$(B_6, C_6)$ and  $(\tA_5, A_5)$ transform as doublets whereas 
$C_4$ and $A_3$ are singlets. 
In particular, for the SL(2,$\Z$) transformation 
\be 
\label{S}
S =\left(\ba{cc} 0 & 1 \\ -1 & 0 \ea \right)\,,
\ee
the field strengths have the required transformation properties. 
Reinstating the dilaton factors one finds that these also behave 
correctly under the transformation $e^{\phi} \rar e^{-\phi}$ associated 
with~(\ref{S}).

\setcounter{equation}{0}
\section{Discussion}

We would like to comment briefly on some of the cases we have not
treated in this paper. For the D4-brane in the type {\II}A theory, the
relevant world-volume field strengths are $F_2$ and $F_3$; since there
is no possible dual to $F_1$ we choose the parameters so that it does
not appear. The tension form is given by
\be
F_5 = \D A_4 - C_5 + x\,B_2\,F_3 - (1{-}x)\,C_3 \, F_2 + x\,C_1\,B_2\,F_2 +
(x{-}\half)\,C_1\,F_2\,F_2 \,,
\ee
and satisfies the Bianchi identity $\D F_5 = R_6 - x\,H_3\, F_3 -
(1{-}x)\,R_4 \, F_2 + (x{-}\half)\,R_2\,F_2\,F_2 = 0$. The duality relations
take the same canonical form as for the other D-branes. For $x=0$ we
recover the usual D4-brane description in terms of $F_2 = \D A_1 -
B_2$, whereas for $x=1$ we obtain a dual description in terms of
$F_3=\D A_2 - C_3 -C_1\,F_2$. In the dual case the action can be
related to the one given in ref.~\cite{Aganagic:1997} by completely
eliminating $F_2$ at the expense of losing manifest gauge
invariance. For the symmetric choice $x=\half$, we obtain an action
which can also be obtained from the M5-brane action given in 
ref.~\cite{Cederwall:1998b} by double dimensional reduction. For the type
{\II}A NS5-brane the relevant world-volume fields are $F_1$, $F_3$ and
$F_5$, while $F_2$ cannot have a dual and is thus not introduced; this
meshes nicely with the fact that the fundamental string 
cannot end on the NS5-brane. The field strength $F_3$ is self-dual,
whereas there is a parameter controlling the dualisation $F_1
\leftrightarrow F_5$. In particular, for a certain choice of the
parameter the action can be related to the direct dimensional
reduction of the action in ref.~\cite{Cederwall:1998b}. By applying the
methods developed in this
paper to the D6-brane, it may be possible---for certain values of the
parameters---to lift the solution to eleven dimensions. Since it is
known that the D6-brane is obtained from the $D{=}11$ 
Kaluza--Klein monopole, one would in this way 
obtain an action for the latter object. The known
action~\cite{Bergshoeff:1997b} for the KK-monopoles gives the standard
D6-brane action upon direct dimensional reduction. 
If it is possible to lift the D6-brane action for other choices of the
parameters one would presumably obtain other formulations of the
action for the KK-monopole in $D=11$. A similar situation holds
for the {\II}A D8-brane, where again it may be possible to lift the solution 
to eleven dimensions for certain values of the parameters and thus make
contact with work on the M9-brane~\cite{Bergshoeff:1998d}. 
In order to treat the last two cases it may be necessary to introduce more 
general parameters than those considered in this paper. It would also 
be of interest to extend the formalism to include the $D=10$ 
KK monopoles. Another issue involves the extension to massive branes
\cite{Bergshoeff:1997c}. We have seen that there is a correspondence
between the possible brane-ending-on-another-brane configurations and 
the restriction to only introducing world-volume fields which have 
duals. Invoking this restriction there seems to be a world-volume field 
for every possible brane-ending-on-another-brane case. Conversely, 
every world-volume field (except for the tension form) can ``arise'' from 
the end of a brane. In addition, one also has the configurations which 
arise from dualisations of the background.
For the type {\II}B branes an interesting outstanding problem concerns
the construction of a manifestly SL(2,$\Z$)-covariant action for the type 
{\II}B $(p,q)$ five-branes. However, the problem of constructing
such an action turns out to be significantly more involved than the 
cases treated in this section; nevertheless some progress can be made 
\cite{WW}. 

There exist formulations of the ten-dimensional $N=2$ supergravity theories
where all the usual bosonic fields (except for the metric) are
``doubled,'' i.e., supplemented with their Poincar\'e duals; see
e.g.\ refs~\cite{Cremmer:1998,Dall'Agata:1998}. In particular, 
the dilaton is supplemented by a nine-form field strength. For each
field allowed in the doubled formulations there
appears to be an associated brane. The question therefore arises if there 
exist branes which couple to the dual of the dilaton (``NS 7-branes''). 
In Hull's general brane scan~\cite{Hull:1997},
which is based on considerations of the supersymmetry algebra of the
background superspace, there appears to be no place for such objects. 
However, in the type {\II}B theory the D7-brane has to transform under
SL(2,$\Z$), since the potential to which it couples does. In particular, 
it is possible to transform it into a brane which couples to the dual of the 
dilaton. It is furthermore known~\cite{Meessen:1998} that in a manifestly 
SL(2,$\Z$)-covariant formulation there must appear a triplet of seven-branes
coupling to the dual eight-form potentials of the three scalars which belong 
to the SL(2,$\R$)/U(1) coset. After gauge fixing there remain two scalars
and two dual eight-form potentials. The branes coupling to these
eight-form potentials are the usual D7-brane and an NS7-brane. 
In the type {\II}A theory, on the other hand, there is only one eight-form 
potential, which in this case should couple to an NS7-brane. 
This object appears to be different from the KK-type seven-brane of
the {\II}A theory discussed in ref.~\cite{Eyras:1998b}.
It would be interesting to see how (or indeed whether) it
fits into the U-duality structure of M-theory
(see e.g.\ ref.~\cite{Obers:1998}).

It would also be desirable to have a more uniform description for the various 
branes within the framework of this paper, along the lines of those available 
for the D-branes both in their original formulation~\cite{Cederwall:1996b,
Cederwall:1996c,Aganagic:1996,Bergshoeff:1996c} 
and in the one with dynamically generated
tension~\cite{Bergshoeff:1998c}. It should also be possible to
construct T-duality rules which relate
actions of the general form considered in this paper. Presumably it is
possible to choose the
parameters so that they are left inert under T-dualisation, as
suggested by the similarity between the D1- and D2-brane actions presented
earlier. A more thorough investigation of the correspondence between
the Bianchi identity for $F_{p+1}$ and $P_+$ which seems to hold in all cases,
might lead to a deeper understanding of $\ka$-symmetry.
Our actions should also find applications in the study of world-volume solitons
\cite{Gibbons:1997,Callan:1997,Gutowski:1998}.

\subsection*{Acknowledgements}

The work of A.W. and N.W. was supported by the European Commission under
contracts FMBICT972021 and FMBICT983302, respectively.

\appendix
\setcounter{equation}{0}
\section{Notation, conventions, and useful formul\ae}
\label{appA}

Throughout the paper we use metrics of signature $(-,+,\cdots,+)$. Moreover,
all our brane actions are written in Einstein frame, i.e., the frame where 
there is no dilaton factor in front of the Einstein--Hilbert term in the 
action for the background supergravity theory.

For superspace forms we use the conventions
\be
\Om_{n} = \mbox{$\frac{1}{n!}$}\D Z^{M_n}\ldots\D Z^{M_1} 
\Om_{M_{1}\ldots M_{n}}
=\mbox{$\frac{1}{n!}$}E^{A_n}\ldots E^{A_1} \Om_{A_1\ldots A_n}\,,
\ee
with the exterior derivative $\D$ as well as the 
interior product $\ip_{V}$, acting from the right, so that
\bea
\D(\Om_m\we\tilde{\Om}_n) &=& \Om_m\we\D\tilde{\Om}_n 
+(-1)^n\D\Om_m\we\tilde{\Om}_n\,, \non\\
\ip_{V}(\Om_m\we\tilde{\Om}_n) &=& \Om_m\we\ip_{V}\tilde{\Om}_n 
+(-1)^n\ip_{V}\Om_m\we\tilde{\Om}_n\,,
\label{extdact}
\eea
where $V$ is an arbitrary super-vector field. (We usually suppress the symbol 
$\we$ when no confusion should arise.)
World-volume forms are defined analogously and hence obey the same rules. 
We do not distinguish notationally between a target-space form $\Om_n$ and 
its pull-back 
to the world-volume, the components of which are given by\footnote{Note that 
$\Om_n$ here denotes a complete $n$-form, not the components of a one-form. 
The distinction should always be clear from the context.}
\be
\Om_{i_1\ldots i_n}=
{E_{i_n}}^{A_n}\ldots{E_{i_1}}^{A_1}\,\Om_{A_1\ldots A_n} :=
\pa_{i_n}Z^{M_n}\,{E_{M_n}}^{A_n}\ldots\pa_{i_1}Z^{M_1}\,{E_{M_1}}^{A_1}\,
\Om_{A_1\ldots A_n}\,.
\ee
The Hodge dual of a world-volume $n$-form is defined by
\be
({*}\Om_n)^{i_{1}\ldots i_{d-n}} = \mbox{$\frac{1}{n!\sqrt{-g}}$}
\ep^{i_1\ldots i_d} \Om_{i_{d-n+1}\ldots i_{d}}\,,
\ee
where $g$ is the determinant of the induced metric 
$g_{ij}\,{=}\,{E_i}^a {E_j}^b\,\eta_{ab}$ and 
$\ep^{i_1\ldots i_d}$ is the totally antisymmetric tensor
density satisfying $\ep^{01\ldots(d-1)}=+1$ and 
$\ep^{i_1{\ldots}i_d}\,\ep^{j_1{\ldots}j_d}=d!\,g\,g^{\mbox{\tiny $\ba{l}
i_1{\ldots}i_d \\ j_1{\ldots}j_d \ea$}}$, $d=p+1$ being the dimension of
the world-volume. 
Here we have defined the tensor
\be
 g^{\mbox{\tiny $\ba{l} i_1{\ldots} i_m \\ j_1 {\ldots} j_m \ea$}} = g^{[i_1
|j_1|}g^{i_2 |j_2|}\ldots g^{i_m] j_m}\,,
\ee
satisfying, in particular, 
\be
g^{\mbox{\tiny $\ba{l} i_1{\ldots i_m}\\ j_1{\ldots}j_m \ea$}}\,
g_{\mbox{\tiny $\ba{l} i_1{\ldots}i_n \\ j_1{\ldots}j_n \ea$}}=
\frac{n!(m-n)!}{m!}\,
g^{\mbox{\tiny $\ba{l} i_1{\ldots}i_{m-n} \\ j_1{\ldots}j_{m-n} \ea$}}\quad
(m\geq n)\,,
\ee
as well as
$g^{\mbox{\tiny$\ba{l} i_1{\ldots}i_m \\ j_1{\ldots}j_m \ea$}}
F_{j_1 \ldots j_m} = F^{i_1 \ldots i_m}$, for any $m$-form $F_m$.

World-volume $\ga$-matrices are defined as the pull-backs 
$\ga_i={E_i}^a\,\Ga_a$, thus inheriting the Clifford algebra 
$\{\ga_i,\ga_j\}=2g_{ij}\id$ from the target-space.
They can be combined into the forms
\be
\ga_n=\mbox{$\frac{1}{n!}$}\,\D\xi^{i_n}\we\ldots\we\D\xi^{i_1}\,
\ga_{i_1\ldots i_n}\,,
\ee
where $\ga_{i_1\ldots i_n}=\ga_{[i_1}\ldots\ga_{i_n]}$ and the 
antisymmetrisation is of weight one.
In particular, we have the world-volume scalar matrix
\be 
\Xi := {*}\ga_{d} = \mbox{$\frac{1}{d!\sqrt{-g}}$}\ep^{i_1\ldots
i_d}\,\ga_{i_1\ldots i_d}\,,
\ee
which satisfies $\Xi^2=(-1)^{\frac{1}{2}d(d-1)+1}\id$.
The latter identity is a special case of the following identity, 
crucial for the $\ka$-symmetry calculations:
\be
\label{general}
\ga_{i_1 \ldots i_m}\ga_{j_1 \ldots j_n} = c^{m,n}_q\,g_{\mbox{\tiny $\ba{l}
[i_1\ldots i_q \\ {[}j_1 \ldots j_q \ea$}}\ga_{\mbox{\tiny $\ba{l}
i_{q+1}\ldots i_m] \\ \phantom{i_{q+1}\ldots i_m}j_{q+1} \ldots j_n]
\ea$}}
\ee 
with the expansion coefficients given by
\be
c^{m,n}_q = \frac{m!}{q!(m-q)!}\frac{n!}{q!(n-q)!}q!(-1)^{(m-q)q +
\frac{1}{2} q(q-1)}\,.
\ee
Finally, the scalar product of two world-volume $n$-forms is defined as 
\be
A_n\scal B_n = \mbox{$\frac{1}{n!}$}A^{i_1\ldots i_n}B_{i_1\ldots i_n}\,.
\ee

\setcounter{equation}{0}
\section{The $D=10$ type {\II} supergravities in superspace}
\label{sugraapp}

The maximally supersymmetric supergravity theories in ten dimensions
were originally formulated in superspace language in refs~\cite{Howe:1984} and 
\cite{Carr:1987} for type {\II}B and type {\II}A, respectively.
However, when dealing with higher-dimensional $p$-branes ($p\geq4$) it
is necessary to use formulations in which the Poincar\'e duals to all
field strengths contained in the respective bosonic sectors are included on 
an equal footing. Since the constraints imposed on the superfields force
the theories on-shell, this is indeed possible to do. 
In fact, one can go even further and consider ``doubled'' 
formulations, in which every bosonic super-field except for the vielbein 
(but including the dilaton) is accompanied by its Poincar\'e-dual field 
(see e.g.\ refs~\cite{Cremmer:1998,Dall'Agata:1998}). 

In its doubled formulation the type {\II}A theory comprises in its
bosonic sector a vielbein ${e_m}^a$, a dilaton $\phi$ and the gauge potentials
$B_2$, $B_6$ and $B_8$, as well as $C_{2k+1}$ ($k=0,\ldots, 4$). 
(In the massive 
theory one also has zero- and ten-form field strengths.) In the superspace 
formulation each of the above fields (as well as the corresponding field 
strengths) becomes the leading component of a superfield. 
In the type {\II}B theory we again have vielbein and dilaton superfields and 
the potentials are $B_2$, $B_6$, $B_8$ and $C_{2k}$ ($k=0,\ldots, 4$). 
The type {\II}B theory is
chiral and has an SO(2) R-symmetry group (U(1) in a complex formulation) 
under which the two Majorana--Weyl spinorial superspace 
coordinates transform as a doublet.

We work with real Majorana spinors. The $\ga$-matrices acting on the spinor 
space are $\ga^a\otimes
\{\id,\ga_{11}\}$ (type {\II}A) and  $\ga^a\otimes \{\id,I,J,K\}$ (type
{\II}B),
where $\ga^a={(\ga^a)^\al}_\bet$ are real; $\ga_{11} = \ga_{0}\ga_1
\cdots\ga_9$ and squares to $\id$; the $2{\times}2$  matrices $I$, $J$ 
and $K$ anticommute pairwise and 
satisfy $I^2 =-\id$, $J^2=\id$, $K^2=\id$, $IJ=K$, $KI=J$ and
$JK=-I$. (Notice in particular the minus-sign in the last relation.)
The matrices $(\ga_{a_1\ldots a_n})_{\al\beta}$ are antisymmetric 
for $n=0,3,4,7,8$ and symmetric for $n=1,2,5,6,9$, while 
$(\ga_{11})_{\al\bet}$ is antisymmetric. Furthermore, the $2{\times}2$
matrix $I$ is antisymmetric, whereas $J$ and $K$ are
symmetric.
 
Spinor indices are raised and lowered with the
antisymmetric charge conjugation matrix $C_{\al\beta}$ and its inverse
$C^{\al\beta}$, according to
the rules $\psi_{\al}= C_{\al\beta}\psi^{\beta}$ and
${M_{\al}}^{\beta} =
C_{\al\la}{M^{\la}}_{\rho}C^{\rho\beta}$. Additional useful information
regarding the 
conventions we use can be found in ref.~\cite{Cederwall:1996c}. 

In the type {\II}B theory, the two physical scalars, $\phi$ and $C_0$, belong
to the coset space 
SL(2,$\R$)/U(1). However, the scalar fields can be made to transform linearly 
under SL(2,$\R$) by combining $\phi$ and $C_0$ into a $2{\times}2$ matrix 
$({\mathcal U}^r,\bar{\mathcal U}^r)$ on which SL(2,$\R$) acts from the
left and U(1) acts locally
from the right. The ${\mathcal U}^r$'s  satisfy  $\frac{i}{2}\ep_{rs}{\mathcal
U}^s\bar{\mathcal U}^r = 1$. By
a suitable fixing of the U(1) gauge symmetry one can remove the 
additional scalar field that has been introduced in the process and regain
the physical scalars. For further details on this construction, see 
ref.~\cite{Howe:1984}.

The on-shell supergravity constraints needed in this paper are\footnote{Here 
we list only the components that enter into our calculations. We use 
the conventions of ref.~\cite{Cederwall:1996c}, where a more complete table of 
constraints can be found (the expression for $H_{a_1\ldots a_5\al\bet}$ 
given above corrects a sign in ref.~\cite{Cederwall:1996c}). 
Finally, $T$ and $\La$ denote the torsion and dilatino superfields, 
respectively.} 
\bea
{\rm {\II}A\&B:}&\hspace{4mm}& {T_{\al\bet}}^{c} =
2i\,(\ga^c)_{\al\bet}\,,\qquad {T_{\al b}}^{c} = 0\,,
\qquad\La_{\al} = \half\pa_{\al}\phi\, \non \\
{\rm {\II}A}: && H_{a\bet\ga}=-2i\,e^{\frac{1}{2}\phi}(\ga_{11}
\ga_a)_{\bet\ga} \non\\
 && H_{ab\ga} = e^{\frac{1}{2}\phi}(\ga_{ab}\ga_{11}\La)_{\ga} \non\\
{\rm {\II}B}: && H_{a\bet\ga}=-2i\,e^{\frac{1}{2}\phi}
(K\ga_a)_{\bet\ga} \non\\
 && H_{ab\ga} = e^{\frac{1}{2}\phi}(\ga_{ab}K\La)_{\ga} \non\\
{\rm {\II}A}: && R_{a_1 \ldots a_{n-2}\al\bet} = 
2i\,e^{\frac{n-5}{4}\phi}(\ga_{a_1\ldots a_{n-2}}
(\ga_{11})^{\frac{n}{2}})_{\al\beta} \non \\
 && R_{a_1 \ldots a_{n-1}\al} = -\mbox{$\frac{n-5}{2}$}
e^{\frac{n-5}{4}\phi}(\ga_{a_1\ldots a_{n-1}}
(-\ga_{11})^{\frac{n}{2}}\La)_{\al} \non \\
{\rm {\II}B}: && R_{a_1 \ldots a_{n-2}\al\bet} = 
2i\,e^{\frac{n-5}{4}\phi}(\ga_{a_1\ldots a_{n-2}}
(K)^{\frac{n-1}{2}}I)_{\al\beta} \non \\
 && R_{a_1 \ldots a_{n-1}\al} = -\mbox{$\frac{n-5}{2}$}
e^{\frac{n-5}{4}\phi}(\ga_{a_1\ldots a_{n-1}}(K)^{\frac{n-1}{2}}I\La)_{\al}
 \non \\
{\rm {\II}B}: && H_{a_1\ldots a_5\al\beta}=-2i\,e^{-\frac{1}{2}\phi}
(\ga_{a_1\ldots a_{5}}K)_{\al\beta}\non \\
 && H_{a_1\ldots a_6\al} = -e^{-\frac{1}{2}\phi}
(\ga_{a_1\ldots a_{6}}K\La)_{\al}
\eea

\setcounter{equation}{0}
\section{$\ka$-symmetry calculations}
\label{kappadetails}

In this appendix we exemplify the calculations performed to verify
$\ka$-invariance of the actions presented in the main part of the paper. 
The methods are similar for all cases, with increasing complexity as the 
number of fields and brane dimensions increase. 
For each case, we also list some identities which follow from the duality 
relations. These identities are crucial for proving $\ka$-invariance and 
should also be useful in applications of our results.
Before delving into the specific cases considered, however, we first
describe some general features of the calculations. 

After inserting a suitably chosen Ansatz for 
$\ka = P_+\ze$ into the $\ka$-variation of the constraint 
$\Ups = 1 + \Phi(\{F_i\}) - ({*}F_{p+1})^2\approx 0$, and utilising
the formula (\ref{general}), the variation takes the generic form
\be
\de_{P_+\ka}\Ups=\sum_{Q,k} \left[ \bLa\, M_Q^{i_1\ldots i_{k}} 
\,\ga_{i_1\ldots i_{k}}
+ \bE_i \,N_Q^{i i_2\ldots i_{k}}\,\ga_{i_2\ldots i_{k}}\right]e^{Q}\,.
\ee
Here $\{e^Q\}=\{\id,\ga_{11}\}$ (type {\II}A) or $\{e^Q\}=\{\id,I,J,K\}$
(type {\II}B). 
Since the $\ga_{i_1\ldots i_{m}}$'s are all linearly independent (for generic 
embeddings), each component has to vanish separately, i.e.
\be
M_{Q}^{[i_1\ldots i_k]}=0\;\;\;
\mbox{(dim 1/2)}\,, 
\quad
N_{Q}^{i[i_2\ldots i_{k}]}=0\;\;\;\mbox{(dim 0)}\,.
\ee
The SO(1,$p$) tensors $N_{Q}^{i[i_2\ldots i_{k}]}$ at dimension 0 can 
be further decomposed into irreducible parts, each of which has to vanish 
separately. We will now give some additional details case by case. 

\subsection*{D2}

The final result of the requirement that all the relevant
$M_{Q}^{[i_1\ldots i_k]}$ and $N_{Q}^{i[i_2\ldots i_{k}]}$ tensors 
 that occur
should vanish is contained in the duality relations 
\bea
{*}F_3\,{*}F_1 &=& F_2 + x\,(F_1\scal F_1)\,F_2 \,,\non \\
-{*}F_3\,{*}F_2 &=& F_1 + (1{-}x)\,(F_2\scal F_2)\,F_1 \,.
\label{D2drapp}
\eea
We will now outline some of the main steps of how to arrive at this
result. 
The projection operator which makes the corresponding $\ka$-variations 
vanish is\footnote{In general it is advantageous to make an Ansatz for 
$P_+$ with arbitrary coefficients which are then determined by the
calculation. To make the presentation more readable we will from the
outset use the correct coefficients.}
\be
2\,{*}F_3\,P_\pm = {*}F_3\id\mp\Big[\Xi - (1{-}x)\,{*}F_1\scal \ga_2\ga_{11} +
x\,{*}F_2\scal\ga_1 \ga_{11} \Big]
\ee
Let us first consider the condition $M_{\sid}^{i}=0$, which reads
$-(3{-}x)\,F^{ik}\,F_k= (3{-}x){*}F^{ik}\,{*}F_k = 0$. 
There are two ways this expression can vanish: 
either $x=3$ or $F^{ik}\,F_k=0$. It turns out that the latter is the case. 
Next we consider the requirement $M_{\ga_{11}}^{[ij]}=0$. 
This condition becomes
\be
\label{g2g11}
\half\,(3{-}x)\left[{*}F_3\,{*}F_1^{ij} - F_{2}^{ij} + x \,{*}(F_1\we
F_2)\,{*}F_1^{ij}\right] = 0\,.
\ee
By using the identity $\ep^{ijk}\,\ep^{lmn}=3!\,g\,g^{\mbox{\tiny $\ba{l}
ijk \\ lmn \ea$}}$ (see appendix \ref{appA}), the last term can be rewritten 
as $-(F_1\scal F_1)\,F_2^{ij} - 2\,F_k\,F^{ik}\,F^j$. 
Using the result of the condition $M_{\sid}^{i}=0$, we see that
(\ref{g2g11}) reduces to the first of the duality relations given in
(\ref{D2drapp}). By similarly analysing the condition $M_{\ga_{11}}^{i}=0$, 
one obtains the second duality relation. For consistency one then has to show
that $F^{ik}\,F_k=0$ follows from the duality relations. This is indeed the 
case, as can be seen by taking the wedge product of the second duality 
relation in (\ref{D2drapp}) with $F_1$. To complete the programme one must 
analyse the remaining components at dimensions $1/2$ and $0$ and show that 
they vanish using the information obtained so far; 
one also has to show that $P_+^2 = P_+$. We will, however, not give the 
details here.

\subsection*{D3}
For the D3-brane case the $\ka$-variation of the constraint 
$\Ups = 1 + \Phi - ({*}F_3)^2 \approx 0$ can be written as
\be
\label{MND3}
\de_{P_+\ka}\Ups=\sum_{Q=0}^3 \left[
\bLa M_Q + \half\bLa M^{i j}_Q \ga_{i j} + 
\bLa \mbox{$\frac{1}{4!}$}M_Q^{i j k l}\ga_{i j k l}
+ \bE_iN_Q^{i j}\ga_{j} + \bE_iN_Q^{i j k l}\ga_{j k l}\right]e^{Q}\,,
\ee
where $\{e^Q\}=\{\id,I,J,K\}$. 
By systematically analysing the components of this expression one can deduce 
the duality relations 
\bea
-{*}F_4\,{*}F_2 &=& \tF_2 - (1{-}x)\,{*}(F_2 \we \tF_2)\,{*}F_2
+ \half\,(1-x)\,{*}(F_2\we F_2)\,{*}\tF_2 - \half\,x\,{*}(\tF_2\we \tF_2)\,
{*}\tF_2 \,,\non\\
{*}F_4\,{*}\tF_2 &=& F_2 - x\,{*}(F_2 \we \tF_2)\,{*}\tF_2 + 
\half\,x\,{*}(\tF_2\we \tF_2)\,{*}F_2 - \half(1{-}x)\,{*}(F_2\we F_2)\,{*}F_2
\,.
\eea
The associated constraint $\Ups \approx 0$ becomes
\bea
&& 1 + (1{-}x)\,F_2\scal F_2 + x\,\tF_2\scal \tF_2 -x\,(1{-}x)\,{*}(F_2\we
 \tF_2)\,{*}(F_2\we\tF_2) \non \\
&&\quad + \half\, x\,(1{-}x)\,{*}(F_2\we
F_2)\,{*}(\tF_2\we\tF_2) -\fourth\,(1{-}x)^{2}\,{*}(F_2\we
F_2)\,{*}(F_2\we F_2) \non \\ 
&&\quad -\fourth\,x^{2}\, 
{*}(\tF_2\we \tF_2)\,{*}(\tF_2\we \tF_2) -({*}F_4)^2 \approx 0\,.
\label{D3constr}
\eea
We would like to stress that the above expressions for the duality
relations and the constraint 
 are the result of requiring the components of (\ref{MND3}) 
to vanish. 
The vanishing of $M_{\sid}^{[ijkl]}$ and the $N_{\sid}^{[ijkl]}$ part of 
$N_{\sid}^{i[jkl]}$ lead to the ubiquitous identity 
\be
\label{D34formid}
F_2\we F_2 + \tF_2 \we\tF_2 = 0 \,.
\ee
Furthermore, modulo this identity, one obtains the first duality relation 
from 
the requirement $M_{J}^{[ij]}=0$ and also from $N_{J}^{[ij]}=0$. Similarly 
from the $M_{K}^{[ij]}=0$ and $N_{K}^{[ij]}=0$ conditions one gets the 
second duality relation. 
The  identity (\ref{D34formid}) can easily be shown to follow from the 
duality relations. As a further example we consider the implications of 
the condition $N_{\sid}^{i j} = 0$. By multiplying the expression for the 
variation with the projection operator one obtains the expression
\bea
\label{Imp}
N_{\sid}^{i j}&=&[1-{*}F_4\,{*}F_4 - 
\al(1{-}x)\,{*}(F_2\we\tF_2){*}(F_2\we\tF_2) +(1{-}x {+}
\al)\,{*}F_4\,{*}(F_2\we\tF_2)]\,g^{ij} \non \\
&&-\,\, (1{-}(x{+}\al))\,{*}{F^{i}}_{l}\,{*}F^{jl} -
(x{+}\al)\,{*}{\tF^{i}}_{l}\,{*}\tF^{jl} +
\al{*}(F_2\we\tF_2)\,{*}{F^{i}}_{l}\,\tF^{jl} \non \\ && +\,\,
\al{*}(F_2\we\tF_2)\,{*}{F^{(i}}_{l}\,\tF^{j)l}- 
{*}F_4\,{*}{F^{(i}}_{l}\,\tF^{j)l}\,.
\eea
This expression consists of three irreducible parts which can be investigated
separately: $N_{\sid}^{[ij]}$, $g^{ij}g_{kl}N_{\sid}^{kl}$ and
$N_{\sid}^{\widetilde{(i j)}}= N_{\sid}^{(i j)} -
\frac{1}{4}\,g_{kl}N_{\sid}^{kl}\,g^{ij}$. 
The antisymmetric part vanishes using the identity ${F^{[i}}_{l}\,\tF^{j]l}=0$
(see below). The trace part is proportional to 
\bea
\label{ga1}
&&1 - {*}F_4\,{*}F_4 + \half\,(1{-}(x{+}\al))\,F_2\scal F_2 + 
\half\,(x{+}\al)\,\tF_2\scal\tF_2 + 
\al\,(x{-}\half)\,{*}(F_2\we\tF_2)\,{*}(F_2\we\tF_2) 
\non \\ &&+\,\, 
(\half{-}x{+}\al)\,\Big[F_2\scal F_2 - x\,{*}(F_2\we \tF_2)\,{*}(F_2\we \tF_2)
 - \half\,{*}(F_2 \we F_2)\,{*}(F_2 \we F_2)\Big]\,.
\eea
Next, we note that it follows from the duality relations that
$x\,K_2\scal F_2 + (1{-}x)\,\tK_2\scal \tF_2 = 0$, which implies 
$F_2\scal F_2 + \tF_2\scal \tF_2 - ({*}(F_2\we\tF_2))^2 - 
({*}(F_2\we F_2))^2 = 0$. Adding $-\half (\al{-}x)$ times this identity to 
the expression (\ref{ga1}) and using (\ref{D34formid}) exactly reproduces 
(\ref{D3constr}).

Finally, to show that the symmetric traceless part is zero it is convenient 
to write ${*}F_4\,{*}{F^{(i}}_{l}\,\tF^{j)l}$ as $(x{+}\al)\,
({*}F_4\,{*}{F^{(i}}_{l})\,\tF^{j)l} + (1{-}(x{+}\al))\,{F^{(i}}_{l}\,
({*}F_4\,\tF^{j)l})$. After inserting the duality relations and using the 
identity ${*}{A^i}_lB^{jl} + {*}{B^i}_lA^{jl} = g^{ij}A\scal{*}B$,
valid for arbitrary $A_{ij}$ and $B_{ij}$, the 
resulting expression for the symmetric traceless part vanishes. 

Some further identities which follow from the duality relations, and which are
instrumental in verifying $\ka$-symmetry, include
${F^{[i}}_{l}\,\tF^{j]l}=0$, ${{*}F^{[i}}_{l}\,\tF^{j]l}=0$,
${F^{[i}}_{l}\,{*}\tF^{j]l}=0$ and ${{*}F^{[i}}_{l}\,{*}\tF^{j]l}=0$,
i.e.\ $F_2$, 
$\tF_2$ ${*}F_2$ and ${*}\tF_2$ all commute when
considered as matrices. This fact, which readily follows from the duality 
relations, means that they can be simultaneously diagonalised. 
Finally we note that one has the freedom to add the identity (\ref{D34formid})
squared to the action without changing the equations of motion; we have fixed
this freedom in a way that yields simple expressions in the limiting
cases $x=0$, $\frac{1}{2}$ and 1.

\subsection*{D5}

The principal difference between the three-brane and the five-brane as far as 
establishing $\ka$-symmetry is concerned, is the increased level of complexity
in the latter case. Otherwise, the basic steps of the computations are
essentially the same and we will therefore try to focus on
the main difficulties. 

First, the condition that $P_+$ (given in (\ref{D5projop})) have the 
properties required of a half-maximal rank projection operator gives three 
independent constraints which must be fulfilled once the duality relations 
have been applied; in addition to an implicit expression for $\Phi$
(similar in structure to the trace-part of the expression given 
in eq.~(\ref{Imp})), these
are the condition $F^{[i|k|}\,{({*}F_4)_k}^{j]}=0$ (i.e., $F_2$ and ${*}F_4$
must commute when considered as matrices) and the identity
\be
\label{D5id}
F_2\we F_2 - {*}F_4\we{*}F_4 - \third\,{*}(F_2\we
F_2\we F_2){*}F_2 = 0\,.
\ee
When turning to the analysis of the conditions for $\ka$-invariance, 
it is very useful to assume the last two properties from the outset and at the 
end show that they follow from the action and the duality relations that 
are derived in the process.

Indeed, all the information needed to determine the action and the 
duality relations is contained in the requirement that 
$\de_{P_+\ze}\Ups\approx0$. 
At dimension 1/2 this gives 16 constraint equations---8 scalar equations 
from the $\Xi$- and $\id$-components plus 8 two-form equations from the 
$\ga_2$ and $\ga_4$ ones. 
At dimension 0, each $\ga_1$-, $\ga_3$- and $\ga_5$-component can be 
decomposed into three SO(1,5)-irreducible tensors which must vanish 
independently. For $\ga_1$ and $\ga_5$ these can be written as the 
antisymmetric, the traceless symmetric and the trace part of a
second-rank tensor. Similarly,
the $\ga_3$-components give rank-four tensors with the symmetries
$N^{[ijkl]}$, $N^{\widetilde{(i[j)}kl]}$ and $g^{i[j}\,{N_m}^{kl]m}$, 
where~$\tilde{\phantom{o}}$~denotes the traceless part.\footnote{Here
$N^{\widetilde{(i[j)}kl]}$ can be further decomposed into self-dual
and anti-self-dual parts. If we write $N^{jkl} :=
N^{\widetilde{(i[j)}kl]}$, these are 
$N^{jkl}_{\pm} = \half(N^{jkl} \pm \frac{\ep^{jklmnp}}{3!\sqrt{-g}}
N_{mnp})$. However, only either $N_+^{jkl} + N_-^{jkl}$ or $N_+^{jkl} 
- N_-^{jkl}$ appear in any
given component constraint.}
Hence, demanding that $(\de_{P_+\ka}\Ups)^{(0)}\approx0$ gives in total 
$3{\times}4{\times}3{=}36$ separate constraint equations, to add to the 16
coming from the corresponding requirement at dimension 1/2.
While some of these 52 constraint equations are more or less trivially 
satisfied, most require some effort to be shown to hold; 
they can roughly be divided into six separate categories: 
trivial ones; those that vanish for purely algebraic reasons; those 
proportional to $[F_2,{*}F_4]$ or the identity (\ref{D5id}); those
that give expressions for the duality 
relations for $F_2$ or $F_4$; and those that give expressions for~$\Phi$. 
(Not all of the constraints are, however, quite as clear-cut, but
rather combinations of the above listed categories.) 

To illustrate the kind of reasoning involved in the analysis, let us consider 
the dimension-0 component\footnote{In this expression and in the remainder of 
the appendix we assume the ``default'' product between rank-two tensors to be 
the matrix product.}
\bea
\de_{P_+\ze}\Ups\big|_{\ga_{(3)}K} &=& 
-2i\,(y+\al)\,{*}F_6\,\bE_i\big[g^{ij}\,F^{kl} +
\sixth {*}{(F_2\we F_2)^i}_m\,{*}{F_2}^{mjkl} \non\\ &&
-\half\,\{({*}F_4\,F_2)^{(ij)}+g^{ij}\,{*}(F_6+y\,F_4\we F_2)\}\,{*}{F_4}^{kl}
\big]\,\ga_{jkl}\,K\,\ze\,.
\eea 
The fully antisymmetric part can be rewritten as 
${N_{\ga_3K}}^{[ijkl]} 
\propto \ep^{ijklmn}\,([F_2,{*}(F_2\we F_2)])_{mn}$
and thus vanishes identically. Using the algebraic identity 
${A^{\widetilde{(i}}}_m\,{*}B^{\widetilde{j)}klm}=
-{B^{\widetilde{(i}}}_m\,{*}A^{\widetilde{j)}klm}$
as well as the identity (\ref{D5id}),
the symmetric, traceless part can be seen to vanish:
\bea
{N_{\ga_3K}}{}^{\widetilde{(i[j)}kl]} &\propto&
{{F_2}^{\widetilde{(i}}}_m\,{*}{F_4}^{|m|\widetilde{j)}}\,{*}{F_4}^{kl]}
+\sixth\,{*}{(F_2\we F_2)^{\widetilde{(i}}}_m\,
{*}{F_2}^{\widetilde{j)}klm} \non \\
&=& \sixth  {{F_2}^{\widetilde{(i}}}_m\,
(F_2\we F_2 {-} {*}F_4\we{*}F_4)^{\widetilde{j)}klm} \non\\
&=& \mbox{$\frac{1}{18}$}\,{*}(F_2\we F_2\we F_2)\, 
{{F_2}^{\widetilde{(i}}}_m\,{*}{F_2}^{\widetilde{j)}klm}= 0\,.
\eea
The remaining component is reminiscent of a duality relation: 
\bea
{(N_{\ga_3K})_j}^{[jkl]} &\propto&
{*}F_6\,{*}{F_4}^{kl} - {F_2}^{kl} + y\,{*}(F_2\we F_4)\,{*}{F_4}^{kl}
+\mbox{$\frac{3}{4}$}\,{({*}F_4\,F_2)_j}^{[j}\,{*}{F_4}^{kl]} \non\\
&& + \fourth\,{*}[F_2\we{*}(F_2\we F_2)]^{kl}\,.
\eea
Indeed, employing the algebraic identity 
${({*}F_4\,F_2)_j}^{[j}\,{*}{F_4}^{kl]}=\third\,
{*}[F_2\we{*}({*}F_4\we{*}F_4)]^{kl}$ (valid when $[F_2,{*}F_4]=0$)
the requirement that ${(N_{\ga_3K})_j}^{[jkl]}=0$ becomes a simplified
version of the duality relation for $F_4$ given in eq.~(\ref{D5dualrels}). 
(Actually, the duality relations are most readily obtained at dimension \half;
more precisely, the components $\ga_4\,I$ and $\ga_2\,K$ give expressions 
for ${*}F_6\,{*}F_2$ and ${*}F_6\,{*}F_4$, respectively.) 

Equipped with expressions for the duality relations one can make use
of the fact that these relations involve the derivatives of the scalar 
functional $\Phi$ in order to determine the action. When doing this,
one has to take into account the further fact that the expressions 
derived from the component analysis are valid only modulo the identity
(\ref{D5id}). The condition that the duality relations both should
give the same $\Phi$ upon integration fixes the expressions and
the end result is the one displayed in~(\ref{D5action}). 

Once the action and the duality relations have been determined, it remains 
to show that the latter imply the property that $F_2$ and ${*}F_4$ commute
as well as the identity (\ref{D5id}). 
The first property is readily proven by taking the commutator of the
Hodge-dual of the first 
duality relation in eq.~(\ref{D5dualrels}) with ${*}F_4$ and
subtracting 
$\frac{1-2y}{1-y}$ times the commutator of the second one with 
$\tF_2$. Then, by using the last of the identities in 
(\ref{D5algids}) below, one is left with an expression proportional to
$[{*}F_4,F_2]$, 
which consequently vanishes. Note that one must use the proper duality
relations for this calculation (i.e., the full expressions given in 
eq.~(\ref{D5dualrels})) in order to be allowed to use the result that 
$[F_2,{*}F_4]=0$ when proving the identity (\ref{D5id}). We will now
turn to this problem. We start by subtracting the exterior product
between ${*}F_4$ and the second duality relation in
eq.~(\ref{D5dualrels}) from the exterior product
between $F_2$ and the Hodge-dual of the first one. One can then show
that one is left with an expression which implies the identity
(\ref{D5id}). In establishing this result the following algebraic identity 
(valid when $[{*}F_4,F_2]=0$) is useful:
\bea
&&{*}(F_2\we F_4)F_2\we{*}F_4 + \half
{*}({*}(F_2\we F_2)\we{*}F_4)\we {*}F_4 +
\half{*}({*}({*}F_4\we{*}F_4)\we F_2)\we F_2 \non \\ && -\,\,\fourth {*}(F_2\we
F_2)\we {*}({*}F_4\we {*}F_4) = 0\,.
\eea

In addition to the identity (\ref{D5id}), a number of purely algebraic 
identities enter in the process of establishing $\ka$-symmetry. Below we 
list a selection of the most frequently occurring ones (here $A$ and $B$ 
denote arbitrary antisymmetric second-rank tensors; note that these
identities are equally useful in the analysis of the NS5-brane case): 
\bea
\third\,{*}(A\we A\we A)\,{*}A - 
\fourth\,{*}(A\we A)\we{*}(A\we A) &=& 0\,, \non\\
{*}[{*}({*}(A\we A)\we A)\we A] + (A\scal A)\,A\we A 
- \third\,{*}(A\we A\we A)\,{*}A &=& 0 \,,\non \\
{A^i}_m\,({*}B)^{jklm}+{B^i}_m\,({*}A)^{jklm}- 3\,g^{i[j}\,{*}(A\we B)^{kl]}
 &=& 0\,, \non \\
{*}({*}(A\we A)\we B)^{[i}{}_{m}B^{j]m} + 
{*}({*}(B\we B)\we A)^{[i}{}_{m}A^{j]m} &=& 0 \,.
\label{D5algids}
\eea
When deriving identities of this kind it is convenient to use the freedom of 
choice of basis to minimise the number of non-vanishing tensor components. 
Also, for the more involved identities, the assistance of a computer algebra 
package can be helpful.

\subsection*{NS5}

The $\ka$-symmetry analysis for the NS5-brane is in many respects similar 
to the D5-brane analysis sketched above. Of course, the fact that there are 
now four instead of two world-volume field strengths does make the analysis
rather more involved and lengthy. 

One quickly finds that the matrices $\tF_2$ and ${*}F_4$ must commute
when the duality relations are satisfied. 
Moreover, combining the condition from the dimension-\half\ component 
$\ga_4$ with that from the totally antisymmetric part of $\ga_4$ at 
dimension 0, leads to the two identities 
\bea
\label{NS5idI}
0 &=& F_0\,F_4 - {*}F_6\,{*}\tF_2 + \half\,{*}F_4\we {*}F_4 - \half\,\tF_2 
\we \tF_2\,, \non\\
0 &=& F_0\,F_4 - {*}F_6\,{*}\tF_2 - \half\,{*}(F_0\,F_4 - \half\,\tF_2\we 
\tF_2) \we {*}(F_0\,F_4 - \half\,\tF_2\we \tF_2)\,.
\eea

Expressions for the duality relations---all valid modulo these 
identities---are obtained from various component constraints.
In particular, the components $\ga_6\,J$, $\ga_4\,I$, $\ga_2\,J$ and $I$
give
\bea
{*}\tF_6\,F_0 &=& {*}F_6 + \sixth\,{*}(\tF_2\we\tF_2\we\tF_2)
+  (1{-}y)\,(F_0)^2\,{*}F_6 - x\,F_0\,{*}(\tF_2\we F_4)\,, \non \\
-{*}\tF_6\,\tF_2 &=& {*}F_4 - (F_0)^2\,{*}F_4  + 
\half\,F_0\,{*}(\tF_2\we\tF_2)+(y{-}2)\,{*}F_6\,F_0\,\tF_2 \non \\ &&
+x\,{*}(\tF_2\we F_4)\,\tF_2 + \half\,{*}[{*}F_4\we{*}(\tF_2\we\tF_2)]
-F_0\,{*}({*}F_4\we{*}F_4) \,, \non \\
{*}\tF_6\,F_4 &=& {*}\tF_2 - y\,{*}F_6\,F_0\,F_4 + 
(1{-}x)\,{*}(\tF_2\we F_4)\,F_4 - F_0\,\tF_2\we{*}F_4 \non \\ &&
+\half\,\tF_2\we{*}(\tF_2\we\tF_2) - \half\,{*}F_6\,\tF_2\we\tF_2 \,, \non \\
-{*}\tF_6\,F_6 &=& {*}F_0 + \third\,F_0\,F_4\we{*}F_4
-\sixth\,\tF_2\we\tF_2\we{*}F_4 
- y\,({*}F_6)^2\,{*}F_0 \non\\ &&
+(x{-}\mbox{$\frac{2}{3}$})\,{*}F_6\,\tF_2\we F_4\,.
\label{NS5simpldr}
\eea
Similar expressions for each of the above duality relations are obtained 
twice at dimension 0. In addition, for other components one obtains a mixture 
of the duality relations as well as various identities following from them, 
which are somewhat intricate to disentangle. 
By inserting the above expressions in trivial identities of the kind 
$0\equiv({*}\tF_6\,F_0)\,F_4-F_0\,({*}\tF_6\,F_4)$, one finds the 
following identities which also enter in the $\ka$-symmetry computations:
\bea
0 &=& {*}F_6\,F_4 - F_0\,{*}\tF_2 - F_0\,{*}(\tF_2\we F_4)\,F_4 + 
F_0\,{*}F_6\,[F_0\,F_4 - \half\,\tF_2\we\tF_2] \non \\ && 
 - \half\,F_0\,\tF_2\we{*}({*}F_4\we{*}F_4)+
\sixth\,{*}(\tF_2\we\tF_2\we\tF_2)\,F_4\,, \non\\
0 &=& \tF_2\we{*}\tF_2 + F_4\we{*}F_4 + {*}(\tF_2\we F_4)\,\tF_2\we F_4
+\mbox{$\frac{3}{4}$}\,\tF_2\we\tF_2\we{*}({*}F_4\we F_4) \non\\ && +
(F_0)^2\,F_4\we{*}F_4 - F_0\,{*}F_4\we\tF_2\we\tF_2 + 
\fourth\,\tF_2\we\tF_2\we{*}(\tF_2\we\tF_2)\,, \non \\
0 &=& F_0\,{*}F_0+F_6\,{*}F_6-\mbox{$\frac{2}{3}$}\,{*}F_6\,F_0\,\tF_2\we F_4
-({*}F_6)^2\,F_0\,{*}F_0 \non \\ && + 
\sixth\,{*}F_6\,\tF_2\we\tF_2\we\tF_2
+\third\,(F_0)^2\,F_4\we{*}F_4 -
\sixth\,F_0\,{*}F_4\we\tF_2\we\tF_2\,.
\eea

The correct expression for $\Phi$ is derived by integrating the expressions
for its functional derivatives obtained from the four duality 
relations~(\ref{NS5simpldr}) via the equations (\ref{NS5dualrels}).
Again, before integrating one must take care to incorporate the freedom 
allowed for in the expressions (\ref{NS5simpldr}) as a consequence of the
identities (\ref{NS5idI}). Most of this freedom is fixed by mutual consistency
requirements and one is left with the expression for $\Phi$ given in 
eq.~(\ref{NS5Phi}). Having obtained an expression for $\Phi$ in this
way, one must show that the identities used to arrive at the result follow 
from the duality relations. In particular, one must derive the condition that 
$\tF_2$ and ${*}F_4$ commute, and in addition the identities~(\ref{NS5idI}). 
In proving the first statement it is convenient to
consider $[{*}K_4,{*}F_4] + [\tF_2,\tK_2]$ and use the algebraic identity
\be
\tF^{[i|m|}{*}(\tF_2\we {*}F_4)_m{}^{j]} - \half{*}(\tF_2\we
 \tF_2)^{[i|m|}{*}F_m{}^{j]} = 0\,.
\ee
Finally, the derivation of~(\ref{NS5idI}) is performed by considering
the trivially satisfied relations
\bea
0 &=& y\,F_0\,{*}K_4+(1{-}x)\,{*}K_6\,\tF_2\,, \non \\
0 &=& \mbox{$\frac{1}{(1{-}y)}$}\,K_0\,{*}F_4+
\mbox{$\frac{1}{(1{-}x)}$}\,{*}F_6\,\tK_2
+\half\,\mbox{$\frac{1}{(1{-}x)}$}
\,{*}K_4\we{*}F_4-\half\,\mbox{$\frac{1}{x}$}\,\tK_2\we\tF_2\,,
\eea
for which the expressions on the right-hand-side can be shown to be linear 
in the identities~(\ref{NS5idI}).

\begingroup\raggedright\endgroup

\end{document}